\documentclass[12pt]{iopart}
\usepackage[caption=false]{subfig}
\usepackage[draft,inline,nomargin,index,marginclue]{fixme} 
\usepackage{iopams}
\usepackage{amssymb}
\usepackage{graphicx}
\usepackage{dcolumn}
\usepackage{bm}
\usepackage{hyperref}
\usepackage{lineno}

\usepackage{rotating}

\begin{document}
\title[]{The atomic damping basis and the collective decay of
  interacting two-level atoms}
\author{W. Alvarez-Giron and P. Barberis-Blostein}
\address{Instituto de Investigaciones en Matem\'aticas Aplicadas y en Sistemas, Universidad Nacional Aut\'onoma de M\'exico, Ciudad
  Universitaria, C.P. 04510, Ciudad de M\'exico, M\'exico.}

\ead{wikkilicht@ciencias.unam.mx}

\begin{abstract}
  We find analytical solutions to the evolution of interacting
  two-level atoms when the master equation is symmetric under the
  permutation of atomic labels. The master equation includes atomic
  independent dissipation. The method to obtain the solutions is: First, we
  use the system symmetries to describe the evolution in an operator
  space whose dimension grows polynomially with the number of
  atoms. Second, we expand the solutions in a basis composed of
  eigenvectors of the dissipative part of the master equation that
  models the independent dissipation of the atoms. This atomic damping
  basis is an atomic analog to the damping basis used for bosonic
  fields~\cite{dampingbases}. The solutions show that the system decays
  as a sum of sub- and super-radiant exponential terms.
\end{abstract}


\submitto{\jpa}
\maketitle

\newcommand{\bra}[1]{\left \langle #1 \right \vert}
\newcommand{\ket}[1]{\left \vert #1 \right \rangle}
\newcommand{\expect}[1]{\left \langle #1 \right\rangle}
\newcommand{\braket}[2]{\left\langle #1 \middle\vert #2 \right\rangle}
\newcommand{\ketbra}[2]{\left \vert #1 \middle \rangle \middle\langle #2 \right \vert}
\newcommand{\adj}[1]{#1 ^\dag}
\newcommand{\braketop}[2]{\left( #1 \middle \vert #2 \right)}
\newcommand{\conm}[1]{\left[ #1 \right]}
\newcommand{\lind}[3]{\hat{#1}^{\mbox{\scriptsize $\begin{array}{cc} #2 \\ #3 \end{array}$}}}
\newcommand{\supop}[3]{\breve{#1}^{\mbox{\scriptsize $\begin{array}{cc} #2 \\ #3 \end{array}$}}}
\newcommand{\vecop}[1]{\left | #1 \right)}
\newcommand{\dualop}[1]{\left( #1 \right|}
\newcommand{\adjop}[1]{\left ( #1 \right | }
\newcommand{\Qform}[3]{#1 ^{\mbox{\scriptsize $\begin{array}{cc} #2 \\ #3 \end{array}$}}}
\newcommand{\subf}[2]{{\small\begin{tabular}[t]{@{}c@{}}
  #1\\#2
  \end{tabular}}
}
\newcommand{\binom}[2]{\left(\begin{array}{c} #1 \\ #2 \end{array}\right)}

\section{Introduction}

Emission processes by interacting quantum emitters exhibit collective
effects~\cite{zoller, ejemplo2, ejemplo3, ejemplo4}. An example of a
quantum emitter is an atom. Atoms can interact with one another through
electromagnetic fields. In free space, collective effects appear when
the distance between the atoms is of the order of the wavelength
associated with the atomic transition~\cite{lehmbergi, agarwal,
  clemens, one-excitation}. When the quantum emitters consist of an
array of two-level atoms near 1D nanowaveguides, the atomic interaction
 is mediated by guided modes. In this case,
the atoms can be far apart and interact with each other, showing
collective effects~\cite{lekien2, solano, arno, chang, tudela}.

Sub- and super- radiance are signatures of collective effects. As an
example, consider two-level atoms in free space. An atom initially in
the excited state will decay exponentially with a rate $\Gamma$.
Something similar happens with several two-level atoms with atomic
distances much larger than the wavelength associated with the atomic
transition. Independently of the initial state, an excitation of a
given atom decays exponentially with rate $\Gamma$, that is, with the
same rate as one atom. If the atoms are close enough, of the order of
the wavelength associated with the atomic transition, the interaction
with each other through the electromagnetic field cannot be neglected.
We focus on the regime where all the atoms are close enough that if a
photon is emitted by an atom, the probability to be reabsorbed by
another is negligible. In this case, the decay of $N$ excited atoms
depends strongly on the initial state~\cite{lehmbergi, agarwal,
  clemens, one-excitation}. For example, if the initial state has one
excitation, but this excitation is shared by all atoms through a
superposition, the atoms decay exponentially. But the decay rate can
be faster or slower than $\Gamma$ depending on the superposition
phases. When interacting atoms decay faster than independent atoms, we
say that the system decay is super-radiant. If the decay is slower we
say it is sub-radiant. In order to have a modification of the decay
rate, coherence between the different quantum emitters is necessary.
Coherence can be generated externally, for example driving the atoms
with a laser. Also, it can appear without using an external source.
In states with more than one excitation without initial coherence,
coherence appears as the number of excited atoms diminishes with time.
Coherence creation is a consequence of atomic interaction through the
electromagnetic field \cite{observables}.

The exponential growth of the number of degrees of freedom with the
number of interacting atoms is one obstacle to studying theoretically
these systems. Analytic solutions for the collective decay have been
found for two \cite{dosatomos, two-atoms} and three atoms
\cite{clemens}. To reduce the exponential complexity for
several atoms one can use the symmetries on the system or study the
thermodynamic limit ($N \to \infty$)~\cite{emary1, emary2}. Also,
there are methods that truncate the Hilbert space of the system, as in
the Matrix Product State Method~\cite{mps}, or constraining the number of
total excitations~\cite{one-excitation}.

If both the master equation, that describes the state evolution, and
the initial condition have symmetries, the evolution of $N$ atoms
takes place in a subspace of operators with a reduced number of
degrees of freedom. An example of this is the super-radiance master
equation without atomic independent decay, where the symmetric Dicke
basis can be used to find solutions \cite{PhysRevA.13.357}. When
atomic independent decay is included in the super-radiance equation,
the symmetric Dicke basis is no longer useful, and one has to use
numerical methods to solve the system for a large number of atoms
\cite{observables}. Nevertheless, the super-radiance equation with
independent atomic decay can be (depending on the position of the
atoms) symmetric under permutation in atomic labels. In this case, the
space of operators acting on the symmetric subspace of the Hilbert
space grows polynomially with the number of atoms. This symmetric
subspace of two-level atoms has been described by Xu
\textit{et al.} \cite{su4} using symmetry transformations of SU(4).
Efficient numerical simulations can be performed using the symmetric
subspace \cite{geremia,gegg2017psiquasp,PhysRevA.98.063815}. The use
of the permutation symmetry has been generalized to $n$-level
systems. Thus, the multiplets of SU($n^2$) have been proposed as a
basis (called \emph{basis of symmetric operators}) for the symmetric
subspace. In addition, the generators of SU($n^2$) (also called
\emph{collective superoperators}) are used to express any linear map
\cite{marduk} on the symmetric subspace.

To get analytical expressions for the evolution of a quantum state,
besides using symmetries to reduce the size of the system space, it is
helpful to find a basis where the evolution of the state takes a
simple form. In the case of master equations that describe
electromagnetic fields in the presence of dissipation (quantum optical
master equations), \emph{damping bases} \cite{dampingbases} have been
useful to obtain analytical solutions. In this approach, the solution
is expanded in a basis given by the eigenvectors of the non-Hermitian
part of the master equation. The method has proven to be helpful
in describing the process of laser cooling~\cite{rv:pablomarc3,
  morigi-cooling}, optomechanical systems~\cite{torres-optomechanical}
and the engineering of quantum states~\cite{morigi-eng}.

The purpose of this paper is to describe the evolution of $N$
interacting two-level atoms using analytical solutions that we obtain
when the master equation is symmetric under the permutation of atomic
indexes. We focus on the case in which the master equation, in the
interaction picture, has only dissipative (independent and collective)
terms (section~\ref{sc:system}). An example of this kind of systems is
a unidimensional equidistant array of atoms near a nanofiber. For this
system, we use the basis of symmetric operators
(section~\ref{sc:subspace}) to find the eigenvectors of the
non-Hermitian part of the master equation describing atomic
independent spontaneous decay (section~\ref{sc:dampingbasis}). These
eigenvectors form a basis, the \emph{atomic damping basis}, that
generalizes for a symmetric system of $N$ atoms the idea of a basis
for the atomic damping of one two-level atom~\cite{dampingbases}. The
atomic damping basis codifies in a convenient way the independent
decay of the atoms. Expanding the solution in this basis we find
analytical expressions using perturbation theory, when the interaction
between the atoms is weak (section \ref{sc:perturbation}), and solving
the differential equations for the coefficients in the expansion in
the general case (section \ref{sc:constriction}).

We focus on the evolution of a symmetric Dicke and a symmetric mixed
state as initial conditions. The symmetric Dicke state exemplifies the
case where the initial state has coherence between its components,
whereas the symmetric mixed state exemplifies the case where there is
no initial coherence between the components. The symmetric mixed state
is particularly interesting since it shows collective behavior
(super-radiance and sub-radiance), which implies that coherence
between the components has been created as the system decays. In an
experiment, comparing the decay dynamics exemplified by these two
cases can be useful to detect if the initial state has coherence
between its components. Analytical expressions are obtained when at
most $M=4$ atoms are excited from a total of $N$.

There are two main results in this work: First, the atomic damping
basis for the symmetric operator subspace that codifies the
dissipation of $N$ independent atoms and, as shown in sections
\ref{sc:perturbation} and \ref{sc:constriction}, is helpful to obtain
solutions of master equations describing interacting atoms. Second,
analytical expressions for the mean number of excited atoms showing the 
decay as a sum of sub- and super-radiant exponential terms.
The weight of each exponential depends on the parameters of the system
and the initial condition. The prediction of super- and sub-radiant
decay in this problem is not new \cite{observables}, but the
analytical expressions we derive are helpful to understand and predict
interesting behavior for different values of the system parameters.
For example, consider the case where the interaction between atoms is
strong and the initial condition consists of a mixed state where $3$
atoms, from a total of $N\gg 3$, are excited, but we do not know which
ones. The evolution of this system creates coherence in such a way
that the sub-radiant decay dominates the evolution at all times and,
as the interactions between the atoms is strong, the decay rate is
very slow.

\section{System}\label{sc:system}
We study $N$ two-level atoms interacting through the electromagnetic
field without external drive. We denote by $|0\rangle^{(\mu)}$ the
ground state and by $|1\rangle^{(\mu)}$ the excited state of atom
$\mu$. We denote by $\mathcal{L}$ the Hilbert space of operators
acting on the system Hilbert space. This space is known as the
\emph{Liouville space} \cite{tarasov}. An operator $O$ is denoted as
$\hat{O}$, except when we refer to a state operator or an element of the
damping basis where we use the rounded ket $\vecop{O}$.


A linear map $B$ acting on the elements of a Liouville space (also
called \emph{superoperator}) is denoted by $\breve{B}$. Thus, the
operator state $\vecop{\tilde{\rho}}\in\mathcal{L}$ satisfies the master
equation
\begin{eqnarray}
  \label{eq:master_equation_complete}
\frac{d}{dt}\vecop{\tilde{\rho}} =  - \frac{i}{\hbar}\conm{\hat{H}_0 + \hat{H}_{\textrm{int}}, \vecop{\tilde{\rho}}} +  \breve{L} \vecop{\tilde{\rho}}\, .
\end{eqnarray}
The unitary part of the evolution is given by the Hamiltonian of the atoms
\begin{eqnarray}\label{h: 0}
\hat{H}_0 = \frac{\hbar \omega_0}{2} \sum_{\mu=1}^N
  \hat{\sigma}_z^{(\mu)} \, ,
\end{eqnarray}
where $\hat{\sigma}_z^{(\mu)} = |1\rangle^{(\mu)}{}^{(\mu)}\langle
1| - |0\rangle^{(\mu)}{}^{(\mu)}\langle
0|$, and the dipolar-dipolar interaction between atoms mediated by the field is 
\cite{lekien}
\begin{eqnarray}\label{h: int}
\hat{H}_{\textrm{int}} = \hbar \sum_{\mu , \, \nu} \Omega_{\mu, \, \nu} \hat{\sigma}_+^{(\mu)} \hat{\sigma}_-^{(\nu)} \, ,
\end{eqnarray}
where $\hat\sigma_-^{(\mu)}=|0\rangle^{(\mu)}{}^{(\mu)}\langle
1|=(\hat\sigma_+^{(\mu)})^\dagger$. 

The non-unitary part of the evolution is described by
$\breve{L}=\breve{L'}_{i}+\breve{L'}_{c}$. The first non-unitary term
is the independent atomic decay modeled by
\begin{eqnarray}
  \breve{L'}_{i} \bullet &=&  \Gamma\sum_{\mu=1}^N  \Big(\hat{\sigma}_{-}^{(\mu)} \bullet \hat{\sigma}_{+}^{(\mu)} - \frac{1}{2} \hat{\sigma}_{+}^{(\mu)}\hat{\sigma}_{-}^{(\mu)}\bullet - \frac{1}{2}
                                \bullet \hat{\sigma}_{+}^{(\mu)}\hat{\sigma}_{-}^{(\mu)}
                                \Big)\, ,
\end{eqnarray}
and the second term is the collective
dissipation \cite{waveguide2}
\begin{eqnarray}
  \breve{L'}_{c} \bullet &=&  \sum_{\mu=1\,\nu=1,\mu\neq\nu}^N
                                \gamma_{\mu\, \nu}\Big(\hat{\sigma}_{-}^{(\nu)} \bullet \hat{\sigma}_{+}^{(\mu)} - \frac{1}{2} \hat{\sigma}_{+}^{(\mu)}\hat{\sigma}_{-}^{(\nu)}\bullet - \frac{1}{2}
                                \bullet \hat{\sigma}_{+}^{(\mu)}\hat{\sigma}_{-}^{(\nu)}
                                \Big)\, . 
\end{eqnarray}
Here $\Gamma$ is the independent atomic spontaneous emission rate and
$\gamma_{\mu\,\nu}$ is proportional to the interaction between the atoms.
When the atoms do not interact with each other, $\gamma_{\mu\,\nu}=0$,
and $\Gamma$ gives the rate at which each atom decays. We will focus on
the case where $\gamma_{\mu\,\nu}=\gamma_c\leq\Gamma$. In this case we
can write $\breve{L}=\breve{L}_{i}+\breve{L}_{c}$ with
\begin{eqnarray}
\label{l10}
  \breve{L}_{i} \bullet &=&  (\Gamma-\gamma_c)\sum_{\mu=1}^N  \Big(\hat{\sigma}_{-}^{(\mu)} \bullet \hat{\sigma}_{+}^{(\mu)} - \frac{1}{2} \hat{\sigma}_{+}^{(\mu)}\hat{\sigma}_{-}^{(\mu)}\bullet - \frac{1}{2}
                                \bullet \hat{\sigma}_{+}^{(\mu)}\hat{\sigma}_{-}^{(\mu)}
                                \Big)\, , 
\end{eqnarray}
and
\begin{eqnarray}  \label{lc}
\breve{L}_{c} \bullet &=& 
\gamma_{c} \bigg[ \hat{J}_- \bullet \hat{J}_+ - \frac{1}{2}
                              \hat{J}_+ \hat{J}_- \bullet -
                              \frac{1}{2}\bullet \hat{J}_+
                              \hat{J}_- \bigg]\, ,
\end{eqnarray}
where the collective atomic operators are
$\hat{J}_{\pm} = \sum_\mu \hat{\sigma}_\pm^{(\mu)}$. This equation is
symmetric under interchange of atomic operator labels.

\begin{figure}
\centering
\includegraphics[width=0.485\textwidth]{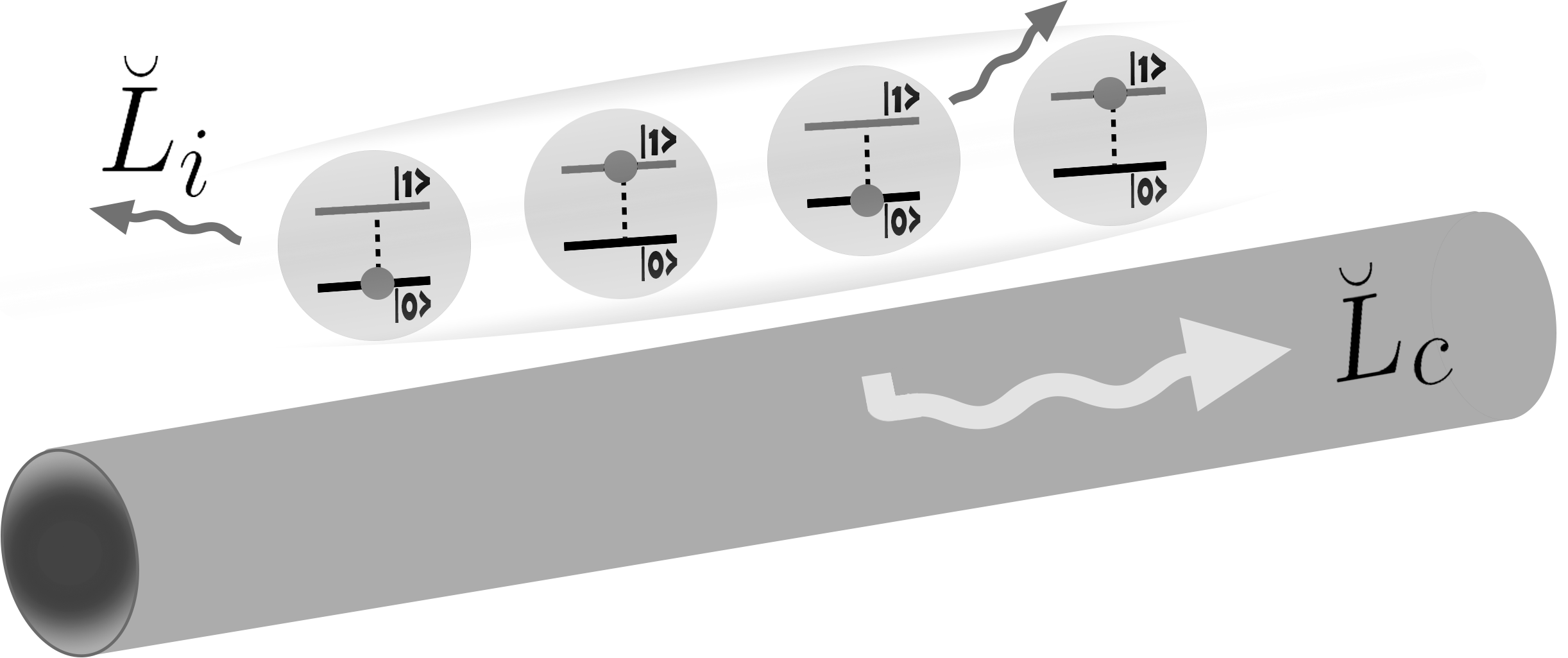}
\caption{Representation of the two dissipation processes for two-level
  atoms in the vicinity of a nanofiber. In the independent
  dissipation, represented by $\breve{L}_{i}$, the atoms emit to free
  space. In the guided collective dissipation, $\breve{L}_c$, the
  atoms can interact between each other even if they are far apart and
  emit along the guided mode of the nanofiber.}
\label{ejemplo}
\end{figure}

Examples where independent and collective dissipation terms, as the
ones we are considering, appear are atoms near a nanowaveguide
\cite{lekien} or inside a leaky cavity \cite{observables}. In the case
of a nanowaveguide (see figure \ref{ejemplo}), the guided
electromagnetic modes introduce long distance interactions between the
atoms. Assuming that the atoms are far apart (the distance between
them is larger than the atomic transition wavelength) the not-guided
(radiative) modes do not introduce dipolar-dipolar interaction terms.
Thus, dipolar coupling between atoms is given only through the guided
modes. This unitary contribution, given by the Hamiltonian \eref{h:
  int}, is a function of atomic position. If we denote by $z_\mu$ the
position of atom $\mu$ along the fiber axial axis,
$\Omega_{\mu, \, \nu}$ is proportional to the sine of
$\beta \left( z_\mu - z_\nu \right)$~\cite{solano}, with $\beta$ the
propagation constant \cite{lekien2}. As we are interested only in
the effect of dissipation terms, we consider interatomic distances
such that $\Omega_{\mu, \, \nu}=0$, making \eref{h: int} zero.

We simplify the master equation~(\ref{eq:master_equation_complete})
using the interaction picture. We define
\begin{eqnarray*}
\vecop{\rho} = \hat{U}^\dag \vecop{\tilde{\rho}} \hat{U} \, 
\end{eqnarray*}
with $\hat{U} = \exp{\left(-i\hat{H}_0 t / \hbar\right)}$. Considering that $\hat{U}^\dag \hat{\sigma}_ \pm^{(\mu)} \hat{U} = \exp (\pm i\omega_0t) \hat{\sigma}_ \pm^{(\mu)} $ and $\hat{H}_{\textrm{int}} = 0$,  Eq. \eref{eq:master_equation_complete} in the interaction picture reads
\begin{eqnarray}\label{eq:master_equation}
\frac{d}{dt} \vecop{\rho} = \breve{L} \vecop{\rho} \, .
\end{eqnarray}

Under these conditions we have two
processes of spontaneous emission. First, we have independent
dissipation where atoms emit photons into radiative modes; second, we
have collective dissipation into the guided modes. Assuming that the
atoms are located in positions with the same coupling to the
nanowaveguide, the master equation for the system is
(\ref{eq:master_equation}) with
$\breve{L}=\breve{L}_{i}+\breve{L}_{c}$.

The formal solution of Eq.~(\ref{eq:master_equation}) is
$\vecop{\rho (t)} = e^{\breve{L} t}\vecop{\rho(0)}$, with
$\vecop{\rho(0)}$ the initial condition. Our goal is to find
$\vecop{\rho(t)}$ by writing the solution as a superposition of the
right eigenvectors of $\breve{L}_{i}$, which satisfy
\begin{equation}\label{eq:right_eigen}
\breve{L}_i \vecop{\alpha} = \lambda_\alpha\, \vecop{\alpha}\, ,
\end{equation}
where $\lambda_\alpha$ is a complex number. Due to the fact that
$\breve{L}_i$ is non-Hermitian, there is no guarantee that a basis of
the Liouville space with the eigenvectors of this linear map exists.
When the set
$\{\vecop{\alpha} : \breve{L}_i \vecop{\alpha} = \alpha
\vecop{\alpha}\}$ forms a basis, it will be called the \emph{atomic
  damping basis}. We denote the dual space of $\mathcal{L}$ as
$\mathcal{L}^*$. Given $\dualop{\alpha} \in \mathcal{L}^*$ we define
the inner product as
\begin{equation*}
  \label{eq:inner_product}
\braketop{\alpha'}{\alpha}\equiv \Tr[\dualop{\alpha'}^\dagger
\vecop{\alpha}] \, .
\end{equation*}
We use the bra-type notation to indicate that $\dualop{\alpha}$ is the
dual of $\vecop{\alpha}$. The elements of $\mathcal{L}^*$ are not
necessarily the Hermitian conjugates of the elements of $\mathcal{L}$.
Therefore, to expand a system state in the damping basis we need to
compute the left eigenvectors
\begin{equation}
\dualop{\alpha} \breve{L}_i = \lambda_\alpha\, 
\dualop{\alpha}\, ,\label{eq:left_eigen}
\end{equation}
where $\dualop{\alpha} \in \mathcal{L}^*$ and satisfy the duality
relation
\begin{equation}\label{eq:inner_product_2}
\braketop{\alpha'}{\alpha}= \delta_{\alpha',
  \alpha}\, .
\end{equation}

With a basis of right and left eigenvectors we can solve the equation
of motion (\ref{eq:master_equation}). For $\breve{L}_{c}=0$ we get
\begin{eqnarray}\label{eq: general-rho}
\vecop{\rho (t)} = \sum_\alpha e^{\lambda_\alpha t} c_\alpha \vecop{\alpha},
\end{eqnarray}
where we used the corresponding left eigenvectors and
Eq.~(\ref{eq:inner_product_2}) to calculate
$c_\alpha = \braketop{\alpha}{\rho(0)}$.

In general the solution will be of the form
\begin{eqnarray}
\vecop{\rho (t)} = \sum_\alpha e^{\lambda_\alpha t} c_\alpha(t)
  \vecop{\alpha}\, ,
\end{eqnarray}
where the coefficients $c_\alpha(t)$ are found using the master
equation.

In the following sections we obtain the damping basis for the symmetric
subspace of $N$ atoms and use it to obtain analytical expressions for
the master equation~(\ref{eq:master_equation}) when
$\breve{L}_c \neq 0$. The damping basis method has been used to obtain
analytical solutions for bosonic systems \cite{dampingbases}. The
program we present here generalizes the idea of the damping basis for
one two-level atom \cite{dampingbases} to a symmetric system of $N$
atoms.

\section{Symmetric subspace}\label{sc:subspace}
The Hilbert space of $N$ two-level atoms is given by the tensor
product
$\mathcal{H}_2\otimes \mathcal{H}_2\otimes\cdots\otimes
\mathcal{H}_2=\mathcal{H}_2^{\otimes N}$, where $\mathcal{H}_2$ is the
space of one two-level atom. The Hilbert space of operators acting on
$\mathcal{H}_2$, the one-atom Liouville space, is denoted by
$\mathcal{L}_4$. For $N$ atoms the Liouville space is
$\mathcal{L}_4^{\otimes N}$ and consists of all the operators that
act on the elements of $\mathcal{H}_2^{\otimes N}$. This operator
space has dimension $4^N$. This exponential growth is reduced when the
system is symmetric under interchange of particle labels. Over the
elements of $\mathcal{L}_4^{\otimes N}$ we can define the permutation
of labels between any pair of particles $i$ and $j$. The operators
invariant under any permutation form the symmetric subspace, denoted
by $\mathcal{S} (\mathcal{L}_{4}^{\otimes N})$, with dimension
$(N+1)(N+2)(N+3)/6$~\cite{su4, Hartmann:2016:GDS:3179439.3179444}.
Because the Lindblad operators \eref{l10} and \eref{lc} remain the
same under any permutation, the evolution of an initial state in
$\mathcal{S} (\mathcal{L}_{4}^{\otimes N})$ under $\breve{L}$ is
constrained to the symmetric subspace.

We introduce a basis for the space
$\mathcal{S} (\mathcal{L}_{4}^{\otimes
  N})$~\cite{Hartmann:2016:GDS:3179439.3179444}. The elements of this basis, called
\emph{basis of symmetric operators}, are \cite{marduk}
\begin{eqnarray}\label{operadoressimetricos}
\fl
\lind{Q}{n_{00}&n_{01}}{n_{10}& n_{11}} &=& \frac{n_{00}! n_{01}! n_{10}! n_{11}!}{N!} \sum_{P} \breve{P} \Big( 
\vecop{00}^{\otimes n_{00}} \vecop{01}^{\otimes n_{01}} 
\vecop{10}^{\otimes n_{10}} \vecop{11}^{\otimes n_{11}}
\Big) \, ,
\end{eqnarray}
where $\vecop{mn}=|m\rangle\langle n| $, $m,n=0,1$, is a basis of
$\mathcal{L}_4$ \cite{tarasov}. For the tensor products in
\eref{operadoressimetricos} we introduce exponents
$n_{mn}=0,1,2,\ldots,N$, which satisfy the constraint
$N = n_{00} + n_{01} + n_{10} + n_{11}$. With this, we denote the
tensor product as
$\vecop{mn}^{\otimes n_{mn}} = \otimes^{n_{mn}}_{k=1} \vecop{mn}$ for
nonzero $n_{mn}$. We use $P$ to indicate some permutation of
$\vecop{00}^{\otimes n_{00}} \vecop{01}^{\otimes n_{01}}
\vecop{10}^{\otimes n_{10}} \vecop{11}^{\otimes n_{11}}$ and
$\breve{P}$ to refer to the superoperator that gives the permutation.
As an example, the symmetric mixed state of two atoms in the ground
state and one atom excited is
\begin{eqnarray*}
\lind{Q}{2&0}{0&1} = \frac{1}{3} \bigg( \vecop{00}\vecop{00}\vecop{11}
                 + \vecop{00}\vecop{11}\vecop{00}
                 +\vecop{11}\vecop{00}\vecop{00} \bigg)\, ,
\end{eqnarray*}
where for simplicity we omit the notation of the tensor product. The
symmetric operators are mutually orthogonal and satisfy
\begin{eqnarray}\label{ortogonal}
\fl
\Tr \left(\lind{Q^\dagger}{n_{00}'&n_{01}'}{n_{10}'& n_{11}'}
\lind{Q}{n_{00}&n_{01}}{n_{10}& n_{11}}
\right) &=& \frac{n_{00}! n_{01}! n_{10}! n_{11}!}{N!}
 \delta_{n_{00}', n_{00}}  \delta_{n_{01}', n_{01}} \delta_{n_{10}', n_{10}}\delta_{n_{11}', n_{11}} \, .
\end{eqnarray}
 Not all the symmetric operators represent physical states, but any operator state in $\mathcal{S} (\mathcal{L}_{4}^{\otimes N})$ can be represented by linear combinations
of these operators.

We introduce the ladder-type superoperators $\supop{A}{ij}{kl}_+$.
These can be written as a sum of $N$ local terms
$\big( \supop{A}{ij}{kl}_+ \big)^{(\mu)}$. On the Liouville space of
each particle we define bosonic superoperators
$\breve{b}_{mn}^{(\mu)}, \, \breve{b}_{mn}^{\dag (\mu)}$ for
$m, \, n = 0, \, 1$. The superoperator $\breve{b}_{mn}^{(\mu)}$
annihilates the operator $\vecop{mn}^{(\mu)}$, while
$\breve{b}_{mn}^{(\mu)\dag}$ creates the operator. With these bosonic
superoperators we define
$\big( \supop{A}{ij}{kl}_+ \big)^{(\mu)} =
\breve{b}_{ij}^{(\mu)\dag}\breve{b}_{kl}^{(\mu)} $, thus the
\emph{collective superoperators} are equal to \cite{marduk}
\begin{eqnarray} \label{coll-supop}
 \supop{A}{ij}{kl}_+ = \sum_{\mu = 1}^N \breve{b}_{ij}^{(\mu)\dag}\breve{b}_{kl}^{(\mu)} \, .
\end{eqnarray}
The superoperators $ \breve{b}$ satisfy the usual rules of commutation $\big[\breve{b}_{ij}^{(\mu)}, \breve{b}_{kl}^{(\mu)\dag} \big] = \delta_{ij, \, kl}$ and $\big[\breve{b}_{ij}^{(\mu)}, \breve{b}_{kl}^{(\mu)} \big] = \big[\breve{b}_{ij}^{(\mu)\dag}, \breve{b}_{kl}^{(\mu)\dag} \big] = 0$. We introduce the bosonic superoperators only as an algebraic support to define the collective superoperators. 

From \eref{coll-supop} we have that $\supop{A}{ij}{kl}_+$ acting on
the left of $\lind{Q}{n_{00} & n_{01}}{n_{10}& n_{11}}$ decrease the
label $n_{kl}$ by one and increase $n_{ij}$ by one, and the resulting
operator is multiplied by $n_{kl}$. For example
\begin{eqnarray} 
\label{a+}
\supop{A}{11}{10}_+ \lind{Q}{n_{00}&n_{01}}{n_{10}&n_{11}} &=& n_{10}\lind{Q}{n_{00}&n_{01}}{n_{10}-1&n_{11}+1}.
\end{eqnarray}
The action of $\supop{A}{ij}{kl}_+$ to the right side of
$\lind{Q^\dagger}{n_{00}&n_{01}}{n_{10}&n_{11}} =
\lind{Q}{n_{00}&n_{10}}{n_{01}&n_{11}}$ is obtained by replacing
$n_{kl} \leftrightarrow n_{ij}$ above. For example
\begin{eqnarray}
\label{ad+}
\lind{Q^\dagger}{n_{00}&n_{01}}{n_{10}&n_{11}} \supop{A}{11}{10}_+  &=& n_{11} \lind{Q^\dagger}{n_{00}&n_{01}}{n_{10}+1&n_{11}-1}.
\end{eqnarray}

\section{Atomic damping basis }\label{sc:dampingbasis}
We write $\breve{L}_i$ in terms of collective
superoperators to solve the eigenvalue problem  \eref{eq:right_eigen} on the symmetric subspace, 
\begin{eqnarray}\label{eq:eq_ladder}
\breve{L}_i \vecop{\rho} &=&  \gamma_{10}\bigg[
\supop{A}{00}{11}_+ - \supop{A}{11}{11}_+ - \frac{1}{2} \bigg( \supop{A}{10}{10}_+ + \supop{A}{01}{01}_+ \bigg)
\bigg] \vecop{\rho}\, , 
\end{eqnarray}
where $\gamma_{10}=\Gamma-\gamma_c$. In \ref{ap1} we show some useful
formulas to obtain the results presented here. As eigenvectors we
propose linear combinations of symmetric operators as
\begin{eqnarray}\label{sym}
\vecop{\rho_{sym}} = \sum_{n_{ij}}
  \Qform{c}{n_{00}&n_{01}}{n_{10}&n_{11}}
                                   \lind{Q}{n_{00}&n_{01}}{n_{10}&n_{11}}\, .
\end{eqnarray}
Using the action of the collective superoperators we find that
\begin{eqnarray*}
\fl
\breve{L}_i \lind{Q}{n_{00}& n_{01}}{n_{10}& n_{11}} &=&
\gamma_{10}\bigg[
n_{11}\lind{Q}{n_{00}+1&n_{01}}{n_{10}&n_{11}-1} -
 n_{11} \lind{Q}{n_{00}&n_{01}}{n_{10}&n_{11}} -
\frac{n_{10}+n_{01}}{2} \lind{Q}{n_{00}&n_{01}}{n_{10}&n_{11}}
\bigg]\,.
\end{eqnarray*}
With the previous equation and substituting~(\ref{sym})
into~(\ref{eq:eq_ladder}) we obtain the recurrence relation
\begin{eqnarray}\label{math:recurrence}
  \Qform{c}{n_{00}-1&n_{01}}{n_{10}&n_{11}+1} &=& \frac{1}{n_{11}+1}\left(
                                     \frac{N}{2} +
                                     \frac{n_{11}-n_{00}}{2} +
                                     \frac{\lambda}{\gamma_{10}} \right)
                                     \Qform{c}{n_{00}&n_{01}}{n_{10}&n_{11}}\,
                                                                      .
\end{eqnarray} 

Solving the recurrence relation we obtain that the eigenvalues are
\begin{eqnarray}
\lambda_{\alpha,\delta} = - \gamma_{10} \bigg[ \frac{N-\alpha}{2} + \delta \bigg],
\end{eqnarray}
where $\alpha = n_{00}+n_{11}$ and $0\leq \delta \leq \alpha$ is an
integer. The right eigenvectors are defined by
{\small
\begin{eqnarray}\label{righteigenvectors}
\fl
\vecop{\alpha, \delta}_n = (-1)^\delta \binom{N}{\alpha} \binom{N-\alpha}{n} 
\binom{\alpha}{\delta} \sum_{n_{11}=0}^\delta (-1)^{n_{11}} \binom{\delta}{n_{11}} 
\lind{Q}{\alpha - n_{11}& N-\alpha-n}{n& n_{11}} \, .
\end{eqnarray}}
For simplicity we denote $n = n_{10}$, which we use to identify the different degenerate eigenvectors.
The eigenvector with $\alpha=N$ and $\delta=0$, $\vecop{N,0}_0$,
represents the ground state. The other eigenvectors do not represent
physical states because their trace is zero. We use them because they
are algebraically easy to manipulate, any symmetrical physical state
can be represented by a superposition of $\vecop{\alpha, \delta}_n$,
and allow us to obtain analytical solutions.

The right eigenvectors \eref{righteigenvectors} are linearly
independent. They are also degenerate because the label $n$ can take
different values, $0 \leq n \leq N-\alpha$. The number of states
$\vecop{\alpha, \delta}_n$ is
\begin{eqnarray*}
\sum_{\alpha=0}^N \sum_{\delta=0}^\alpha (N-\alpha + 1) =
  \frac{(N+1)(N+2)(N+3)}{6}\, .
\end{eqnarray*}
As the number of eigenvectors matches the dimension of
$\mathcal{S}(\mathcal{L}_4^{\otimes N})$, the operators \eref{righteigenvectors} form a basis for
the symmetric subspace.

We need to know the left eigenvectors to find the coefficients necessary to expand any operator in the 
damping basis. Applying the
previous method for the eigenvalue problem
\begin{equation}
\dualop{\rho_{sym}}\breve{L}_i = \lambda
\dualop{\rho_{sym}}\, ,\label{eq:left_eigen_system}
\end{equation}
we get the left eigenvectors
\begin{eqnarray}
\label{lefteigenvectors}
_{n}\dualop{\alpha, \delta} &=&  \sum_{n_{11} = \delta}^{\alpha} \binom{\alpha-\delta}{\alpha- n_{11}} \lind{Q}{\alpha -n_{11}& N-\alpha -n}{n& n_{11}}.
\end{eqnarray}
The left eigenvectors are dual to the right eigenvectors, i.e. they
satisfy
\begin{displaymath}
_{n'}\braketop{\alpha', \delta'}{\alpha, \delta}_n = \delta_{\alpha',
  \alpha} \delta_{\delta', \delta} \delta_{n',n}\, .
\end{displaymath}
Using this relation we can express any symmetric operator in terms of
the atomic damping basis as
\begin{equation}\label{eq:Quads}
\lind{Q}{\alpha-n_{11} &N-\alpha -n }{n& n_{11}} =
\sum_{k=0}^{n_{11}} c_k \vecop{\alpha, k}_n\, ,
\end{equation}
with
\begin{eqnarray}\label{qinverso}
c_k
=
 \left[ 
\binom{\alpha}{n_{11}}\binom{N-\alpha}{n}\binom{N}{\alpha}
\right]^{-1} 
\binom{\alpha-k}{\alpha - n_{11}}\, .
\end{eqnarray}
The solution of $ d\vecop{\rho} /dt = \breve{L}_i \vecop{\rho}$, with
initial condition
$\vecop{\rho(0)} = \lind{Q}{\alpha-\delta &N-\alpha -n }{n& \delta}$,
is
\begin{eqnarray}
\vecop{\rho(t)} &=& e ^{-\frac{N-\alpha}{2} \gamma_{10} t } 
\frac{n!(N-\alpha-n)!}{N!}
\sum_{i=0}^{\delta} d_{i}(t)\lind{Q}{\alpha -i& N-\alpha-n}{n& i},
\end{eqnarray}
where
\begin{eqnarray}
d_i(t) = \sum_{j = i}^\delta (-1)^i c_j 
\binom{j}{i} 
i!\left(\alpha - i \right)!e^{-j\gamma_{10} t}.
\end{eqnarray}

\section{Perturbation of the atomic damping
  basis}\label{sc:perturbation}
A powerful use of the damping basis is to find analytical
solutions to the master equation~(\ref{eq:master_equation}) when
$\gamma_c\ll\Gamma$. Linear superpositions of the damping basis
are the solutions to Eq.~(\ref{eq:master_equation}) when $\gamma_c=0$.
Using perturbation theory in Liouville space, we can find the
eigenvalues and eigenvectors of $\breve{L}$ as a perturbation of the
damping basis. Using the perturbed eigenvectors and eigenvalues
we can find analytical expressions to the quantum state evolution.
\subsection{Perturbation theory in Liouville space}

We follow \cite{sakurai} to derive a perturbation method for the
eigenvalues and eigenvectors of $\breve{L}$ in the degenerate case. We
denote by $\vecop{\kappa}$ and $\dualop{\kappa}$ the right and left
eigenvectors of $\breve{L}_0$, respectively. We denote by
$\lambda^{(0)}_\kappa$ its eigenvalues. We want to solve the
eigenvalue equation
\begin{eqnarray}
\left( \breve{L}_0 
+ \gamma 
\breve{L}_{\textrm{pert}} 
\right)
\vecop{\phi_\kappa} = \lambda_\kappa \vecop{\phi_\kappa}\, ,
\end{eqnarray}
where $\gamma \breve{L}_{\textrm{pert}}$ is the perturbation, with $\gamma \ll 1$ giving its strength.

Let us assume that the eigenvectors with eigenvalue $\lambda_\kappa ^{(0)}$
are degenerate and consider the subspace $\{\vecop{\kappa}\}$ created
by them. In this space we can form the projectors
\begin{eqnarray*}
\mathbb{Q}_{\kappa}= \sum_{\kappa}\vecop{\kappa}\dualop{\kappa}, \qquad \mathbb{P}_{\kappa} = \mathbb{I} - \mathbb{Q}_{\kappa},
\end{eqnarray*}
which are idempotent, commute with each other and commute with $\breve{L}_{0}$. 
If we use these projectors in the eigenvalues equation we obtain
\begin{eqnarray}\label{ec:pertestados}
\breve{L}_{0} \left(\mathbb{Q}_{\kappa} + \mathbb{P}_{\kappa} \right) \vecop{\phi_{\kappa}} + \gamma \breve{L}_{\textrm{pert}} \left(\mathbb{Q}_{\kappa} + \mathbb{P}_{\kappa} \right)
 \vecop{\phi_{\kappa}} =
   \lambda_{\kappa} \left( \mathbb{Q}_{\kappa} + \mathbb{P}_{\kappa} \right) \vecop{\phi_{\kappa}}.
\end{eqnarray}

We multiply~\Eref{ec:pertestados} by $\mathbb{P}_{\kappa}$ and
solve for $\mathbb{P}_{\kappa} \vecop{\phi_{\kappa}}$. Then we
multiply~(\ref{ec:pertestados}) by $\mathbb{Q}_{\kappa}$ and introduce
the expression for $\mathbb{P}_{\kappa} \vecop{\phi_{\kappa}}$ to obtain
{\small
\begin{eqnarray*}
\fl
\left(\lambda_{\kappa} - \breve{L}_{0} - \gamma \mathbb{Q}_{\kappa}\breve{L}_{\textrm{pert}} \mathbb{Q}_{\kappa}\right) \mathbb{Q}_{\kappa}\vecop{\phi_{\kappa}}
= 
\gamma^2 \mathbb{Q}_{\kappa} \breve{L}_{\textrm{pert}}
  \left(\lambda_{\kappa} - \breve{L}_{0} - \gamma
  \mathbb{P}_{\kappa}\breve{L}_{\textrm{pert}}
  \mathbb{P}_{\kappa}\right)^{-1}  \mathbb{P}_{\kappa}
  \breve{L}_{\textrm{pert}} \mathbb{Q}_{\kappa}\vecop{\phi_{\kappa}}\, .
\end{eqnarray*}}
At first order in $\gamma$ we obtain
\begin{eqnarray}\label{perturbacion}
\left(\lambda_{\kappa} - \lambda_{\kappa}^{(0)} - \gamma \mathbb{Q}_{\kappa}\breve{L}_{\textrm{pert}} \mathbb{Q}_{\kappa}\right) \mathbb{Q}_{\kappa} \vecop{\phi_{\kappa}}
= 0\, .
\end{eqnarray}

Let $\vecop{a} \equiv \mathbb{Q}_{\kappa} \vecop{\phi_{\kappa}}$, which
satisfies
$\mathbb{Q}_{\kappa}\breve{L}_{\textrm{pert}} \mathbb{Q}_{\kappa}
\vecop{a} = a \vecop{a}$, with $a$ a scalar. Then $\vecop{a}$ is an
eigenvector of \Eref{perturbacion} with eigenvalue
$\lambda_{\kappa} = \lambda_{\kappa}^{(0)} + \gamma a$. To find the
left eigenvectors $\dualop{a}$ we construct the matrix \textbf{A} with
the right eigenvectors as columns. The left eigenvectors are the rows
of \textbf{A}$^{-1}$ \cite{vector1, vector2}.

\subsection{Perturbation with $\breve{L}_c$}  

We apply perturbation theory to \Eref{eq:master_equation} when
$1 \gg \gamma \equiv \gamma_{c}/\Gamma$. Physically this means
that most of the photons are dissipated to free space and a small
fraction are emitted into the guided mode. Using the identities in 
\ref{ap1} we write $\breve{L}_c$ as
\begin{eqnarray} \label{supopcoll}
\fl
\breve{L}_c \vecop{\hat{\rho}} = 
 \gamma_{c}\bigg[
\supop{A}{00}{10}_+\supop{A}{10}{11}_+ + \supop{A}{00}{10}_+\supop{A}{00}{01}_+ + 
\supop{A}{01}{11}_+\supop{A}{10}{11}_+ + \supop{A}{01}{11}_+\supop{A}{00}{01}_+ 
- \frac{1}{2} \bigg( \supop{A}{10}{00}_+\supop{A}{00}{10}_+ 
\nonumber \\ 
+ \supop{A}{10}{00}_+
\supop{A}{01}{11}_+ + \supop{A}{11}{01}_+\supop{A}{00}{10}_+ +
\supop{A}{11}{01}_+ \supop{A}{01}{11}_+ 
 + \supop{A}{11}{10}_+\supop{A}{10}{11}_+ + \supop{A}{11}{10}_+\supop{A}{00}{01}_+ 
 \nonumber \\ 
 + \supop{A}{01}{00}_+\supop{A}{10}{11}_+ +
 \supop{A}{01}{00}_+\supop{A}{00}{01}_+    
\bigg) \bigg]\vecop{\rho}\, .
\end{eqnarray}

Given $\alpha, \delta$, we have $N-\alpha + 1$ eigenvectors. In
addition, different values of labels $\alpha, \delta$ can have the
same eigenvalue. Specifically
$\lambda_{\alpha, \delta} = \lambda_{\alpha -2m, \delta -m}$ for
$m \in \mathbb{Z}$, $0\leq \alpha -2m \leq N$ and
$0\leq \delta - m \leq \alpha - 2m$. We denote the set of labels that
meet the above criteria as $\{\alpha, \delta \}$.

The projector for the degenerate subspace with eigenvalue
$\lambda_{\alpha, \delta}$ is
$\mathbb{Q}_{\alpha,\delta} = \sum_{\{\alpha, \delta\}, n}
\vecop{\alpha, \delta}_n\, _n\dualop{\alpha, \delta}$), and for the collective term we obtain 
{\small
\begin{eqnarray}\label{eq:pert_matrix}
\fl
\mathbb{Q}_{\alpha, \delta} \breve{L}_{c} \mathbb{Q}_{\alpha, \delta} =-\gamma_c \sum_{\{\alpha, \delta\}} \sum_{n=0}^{N - \alpha}\bigg[
(\delta +1)(\alpha-\delta+1)
\vecop{\alpha +2, \delta+1}_{n-1} \, _n\dualop{\alpha, \delta}
+  \Bigg\{
\frac{(N-\alpha)(\alpha +1)}{2}  
\nonumber \\ 
+ \delta \Bigg\}
\vecop{\alpha, \delta}_{n} \, _n\dualop{\alpha, \delta}
+ (n+1)(N-\alpha-n+
1)
\vecop{\alpha -2, \delta  -1}_{n+1} \,_n\dualop{\alpha, \delta}\bigg]
                \, .
\end{eqnarray}}
This result is the matrix that we must diagonalize to
obtain the damping basis corrections. We just need to identify each
subspace generated by the eigenvalues, and evaluate
\Eref{eq:pert_matrix}. The physical systems for which this result
is valid include atoms coupled to nanofibers where values of
$\gamma \sim 0.05$ have been reported \cite{nanofibras}.

\subsubsection{Example: $N$ atoms with up to 4 excitations}\label{sec:3-exc} 

We use time-independent perturbation theory to calculate the evolution
of $N$ atoms under the master equation \eref{eq:master_equation}, when
the maximum number of initially excited atoms is $M=4$. Because there
is no external drive in the master equation, the evolution cannot
exceed $M$ excited atoms. Therefore, in \Eref{eq:pert_matrix} we
consider as zero all those eigenvectors that have less than $N-4$
atoms in the ground state. Under these approximations we restrict the
evolution of the operator state to a subspace of dimension
$(M+1)(M+2)(M+3)/6$. In \ref{ap2} we show the eigenvalues and the
right eigenvectors of $\breve{L}_{i}$ for the subspace limited to
$M=3$ excitations, and the perturbation of these eigenvalues and
eigenvectors due $\breve{L}_c$.

We use the perturbed eigenvectors to solve the evolution of two
initial conditions: a symmetric mixed state and a symmetric Dicke
state.

The symmetric mixed state of $M$ excitations is
\begin{eqnarray}\label{eq:initial_mixed}
\vecop{\rho_{\textrm{mixed}}^{(M)}} = \lind{Q}{N-M&0}{0& M}\, .
\end{eqnarray}
This consists of $M$ excited atoms out of a total of $N$ atoms, but we
do not know which ones. Using \Eref{eq:Quads} we can write
$\vecop{\rho_{\textrm{mixed}}^{(M)}}$ in the damping basis. In order
to calculate the time evolution we use the perturbed basis. This basis
is shown in \ref{ap2} for the case $M=3$ and the analytical expression
for the time evolution of $\vecop{\rho_{\textrm{mixed}}^{(M)}}$ is
given by \Eref{soluciontresexc} and \Eref{soluciontresexc-mixed}.

We are interested in the mean number of excited atoms $P(t)$. Any
observable of $\mathcal{S}(\mathcal{L}_4^{\otimes N})$ can be written
in terms of collective superoperators. In particular we have in the
interaction picture,
\begin{eqnarray}
\fl
P(t) = \left\langle \sum_\mu \hat{\sigma}_+^{(\mu)}(t) \hat{\sigma}_-^{(\mu)}(t)
  \right\rangle = \left\langle \sum_\mu \hat{\sigma}_+^{(\mu)} \hat{\sigma}_-^{(\mu)}
  \right\rangle  = \Tr \left[ \left( \supop{A}{10}{10}_+ +
  \supop{A}{11}{11}_+ \right) \vecop{\rho(t)} \right] \, ,
\end{eqnarray}
where we have used that $\hat{\sigma}^{(\mu)}_\pm (t) = \exp(\pm i\omega_0 t) \hat{\sigma}^{(\mu)}_\pm$. 
For $M=1, \, 2, \, 3, \, 4$ we obtain
{\small
\begin{eqnarray}\label{ec:irm}
P_{ \textrm{mixed}}^{(M)} (t) &=& M \left[ \frac{N-1}{N}e^{- \left(\Gamma - \gamma_c \right) t} + \frac{1}{N}e^{-\left[\Gamma + (N-1)\gamma_c \right] t} \right] \, .
\end{eqnarray}}
When atoms do not interact with each other ($\gamma_c=0$), the mean
number of excited atoms decays as $M e^{-\Gamma t}$. When atoms
interact with each other, the mean number of excited atoms is composed
of a sub-radiant part, that decays with rate $\Gamma - \gamma_c$, and a
super-radiant part, that decays with rate $\Gamma + (N-1)\gamma_c $,
which increases as the number of atoms increases. The initial state
does not have coherence between different atoms. As the system
evolves, the interaction between atoms through the field creates the
coherence that explains the sub- and super-radiant behavior. A similar
effect happens to spatially close atoms in free space \cite{clemens}.
When $N\gg 1$, the sub-radiant contribution to
$P_{\textrm{mixed}}^{(M)}$ dominates the evolution (the weight of the
term is $(N-1)/N$) with respect to the super-radiant contribution (a
contribution of $1/N$ in the evolution). A signature that the
sub-radiant part dominates the evolution is that
$P_{\textrm{mixed}}^{(M)}(t) \geq M\,e^{-\Gamma t}$. Note that the
relative contribution of the sub-radiant part with respect to the
super-radiant part does not depends on $M$, only on $N$. If $N>1$ we
always have a super- and sub- radiant contribution, this can be
explained by the fact that the initial state can be written as a
superposition of a sub- and super- radiant states.

The \Eref{ec:irm} has a very simple form. We use the quantum
trajectory formalism \cite{lb:carmichael} to explain it. In this
formalism, the effect of atomic decay is modeled as a quantum state
trajectory given by a series of random quantum jumps, and a
non-Hermitian evolution between them. The expectation value of an
observable is obtained as a weighted sum of the expectation value for
each trajectory. We assume that the initial state is
$|1\rangle^{(1)}|1\rangle^{(2)}$. Consider the trajectory in which the
first atom decays through the independent dissipation channel, so that
the state collapses to $|0\rangle^{(1)}|1\rangle^{(2)}$, which can be
written as
\begin{equation}
\fl
  \label{eq:trajectory}
  |0\rangle^{(1)}|1\rangle^{(2)}=\frac{1}{2}\Big(|1\rangle^{(1)}|0\rangle^{(2)}+|0\rangle^{(1)}|1\rangle^{(2)}\Big)-\frac{1}{2}\Big(|1\rangle^{(1)}|0\rangle^{(2)}-|0\rangle^{(1)}|1\rangle^{(2)}\Big)\, .
\end{equation}
The first term in the sum is a symmetric Dicke state with one
excitation, which is a super-radiant state that decays as
$e^{-\left[\Gamma + \gamma_c \right] t} $; the second term is a
sub-radiant state that decays as $ e^{-(\Gamma - \gamma_c) t} $. A
similar analysis can be done when the second atom decays. Now consider
the initial state with three excitations,
$|1\rangle^{(1)}|1\rangle^{(2)}|1\rangle^{(3)}$, and assume that the
first atom independently decays. We then have
\begin{eqnarray}
  \label{eq:3ndecaytrajectory}
|0\rangle^{(1)}|1\rangle^{(2)}|1\rangle^{(3)}&=&\frac{1}{3}\Big(|0\rangle^{(1)}|1\rangle^{2)}|1\rangle^{(3)}+|1\rangle^{(1)}|0\rangle^{2)}|1\rangle^{(3)}+|1\rangle^{(1)}|1\rangle^{2)}|0\rangle^{(3)}\Big)\nonumber\\&&+\frac{1}{3}\Big(|0\rangle^{(1)}|1\rangle^{2)}|1\rangle^{(3)}-|1\rangle^{(1)}|0\rangle^{2)}|1\rangle^{(3)}\Big)\nonumber\\&&+\frac{1}{3}\Big(|0\rangle^{(1)}|1\rangle^{2)}|1\rangle^{(3)}-|1\rangle^{(1)}|1\rangle^{2)}|0\rangle^{(3)}\Big)\, .
\end{eqnarray}
The first term in the sum is a symmetric Dicke state with two
excitations, which is a super-radiant state that decays as
$e^{-\left[\Gamma + 2\gamma_c \right] t} $; the second and third terms
are sub-radiant states, both decay as $ e^{-(\Gamma - \gamma_c) t} $.
Similar results can be obtained for other cases. The simple form of
Eq.~(\ref{ec:irm}) is a consequence of the fact that the quantum
trajectory created by the independent decay process, which is dominant
for $\Gamma\gg\gamma_c$, can be written as a superposition of states
that decay with two different rates: a sub-radiant decay and a
super-radiant decay.


The symmetric Dicke state with $M$ excitations, 
\begin{eqnarray}\label{eq:initial_dicke}
\vecop{\rho_{\textrm{dicke}}^{(M)}} &=&  \sum_{0=i+j\leq M} \frac{M!}{i!j!(M-i-j)!}\lind{Q}{N-M&i}{j&M-i-j}\, ,\nonumber\\
\end{eqnarray}
represents a pure state with $M$ excitations shared by $N$ atoms. 
Using \Eref{eq:Quads} we can write $\vecop{\rho_{\textrm{dicke}}^{(M)}}$
in the damping basis and use the perturbed damping basis
to calculate the evolution. The result for $M=3$ is shown by 
\Eref{soluciontresexc} and \Eref{soluciontresexc-dicke}.

The mean number of excited atoms for $M=1,\,2,\,3,\,4$ is  {
\begin{eqnarray}
P_{ \textrm{dicke}}^{(M)}(t) &=& M \Bigg[\frac{N-M+1}{N}e^{- \left[\Gamma + (N-1)\gamma_c \right]t} +
 \frac{M-1}{N}e^{-\left(\Gamma  - \gamma_c \right) t} \Bigg] \,. \label{ec:ird}
\end{eqnarray}}
Similar to the case where the initial condition is a symmetric mixed
state with $M$ excited atoms, the decay  has a sub- and super- radiant contribution. 
Differently to that case,  $M$ determines the contribution of the
sub- and super-radiant terms. When
$M=1$ there is no sub-radiant term: the initial state is
super-radiant,
once the atom decays it reaches the ground state.  For
the symmetric Dicke state, $P_{\textrm{dicke}}^{(M)}(t) \leq
M\,e^{-\Gamma t}$ for short times. The super-radiant decay
dominates the initial evolution. For large times $P_{ \textrm{dicke}}^{(M)}(t) \geq
M\,e^{-\Gamma t}$ and the sub-radiant decay dominates the evolution.

\section{Solutions without perturbation}\label{sc:constriction}
In this section we consider the master equation
\eref{eq:master_equation} without assuming any constraint on the
values of $\Gamma$ and $\gamma_c$. Systems where perturbation theory
is no longer valid include atoms coupled to a photonic crystal
\cite{cristales}, and quantum dots coupled to waveguides
($\gamma_c/[\Gamma-\gamma_c] \gg 1$) \cite{quantumdots}. The general
idea is to solve the master equation by writing $\breve{L}$ as a matrix
using the damping basis, and finding the eigenvectors and eigenvalues.
We consider a subspace of $\mathcal{S} (\mathcal{L}_{4}^{\otimes N})$
given by the span of
$\mathcal{A}_3 = \{\vecop{N,0}_0, \vecop{N,1}_0, \vecop{N-2,0}_1,
\vecop{N,2}_0, \vecop{N-2,1}_1, $
$ \vecop{N-4,0}_2, \vecop{N,3}_0, \vecop{N-2,2}_1,\vecop{N-4,1}_2,
\vecop{N-6,0}_3 \}$, where the evolution for the symmetric Dicke and
mixed states with $M=1,\, 2, \,3$ occurs. We solve the master
equation~\eref{eq:master_equation} by diagonalizing the matrix
representation of $\breve{L}_{i} + \breve{L}_c$ in this subspace. The
matrix is shown in \ref{ap3}.


In the perturbative
case we only need to apply formula \eref{eq:pert_matrix} to obtain the
matrix, instead of calculating all the mappings. We use the
eigenvectors of the matrix shown in \ref{ap3} to obtain the state
evolution without perturbation.

The general solution for the quantum state with up to $M=3$ initial excitations
has the form
{\small
\begin{eqnarray}\label{estado3}
\fl
\vecop{\rho(t)} = 
\alpha_1(t) \vecop{N,0}_0 + \alpha_2(t)\vecop{N,1}_0 + \alpha_3(t) \vecop{N-2,0}_1 + 
 \alpha_4(t) \vecop{N,2}_0 + 
 \alpha_5(t) \vecop{N-2,1}_1 +
 \nonumber \\ \alpha_6(t) \vecop{N-4,0}_2   
 + \alpha_7(t)\vecop{N,3}_0  +  \alpha_8(t)\vecop{N-2,2}_1 + 
 \alpha_9(t)\vecop{N-4,1}_2   +
 \nonumber \\
\alpha_{10}(t)\vecop{N-6,0}_3\, . 
\end{eqnarray}}
The analytical expressions for the functions $\alpha_i(t)$ are too large
to be included in the text, but they can be obtained using a Computer
Algebra System (CAS).

\begin{figure}
\centering
\includegraphics[width=0.485\textwidth]{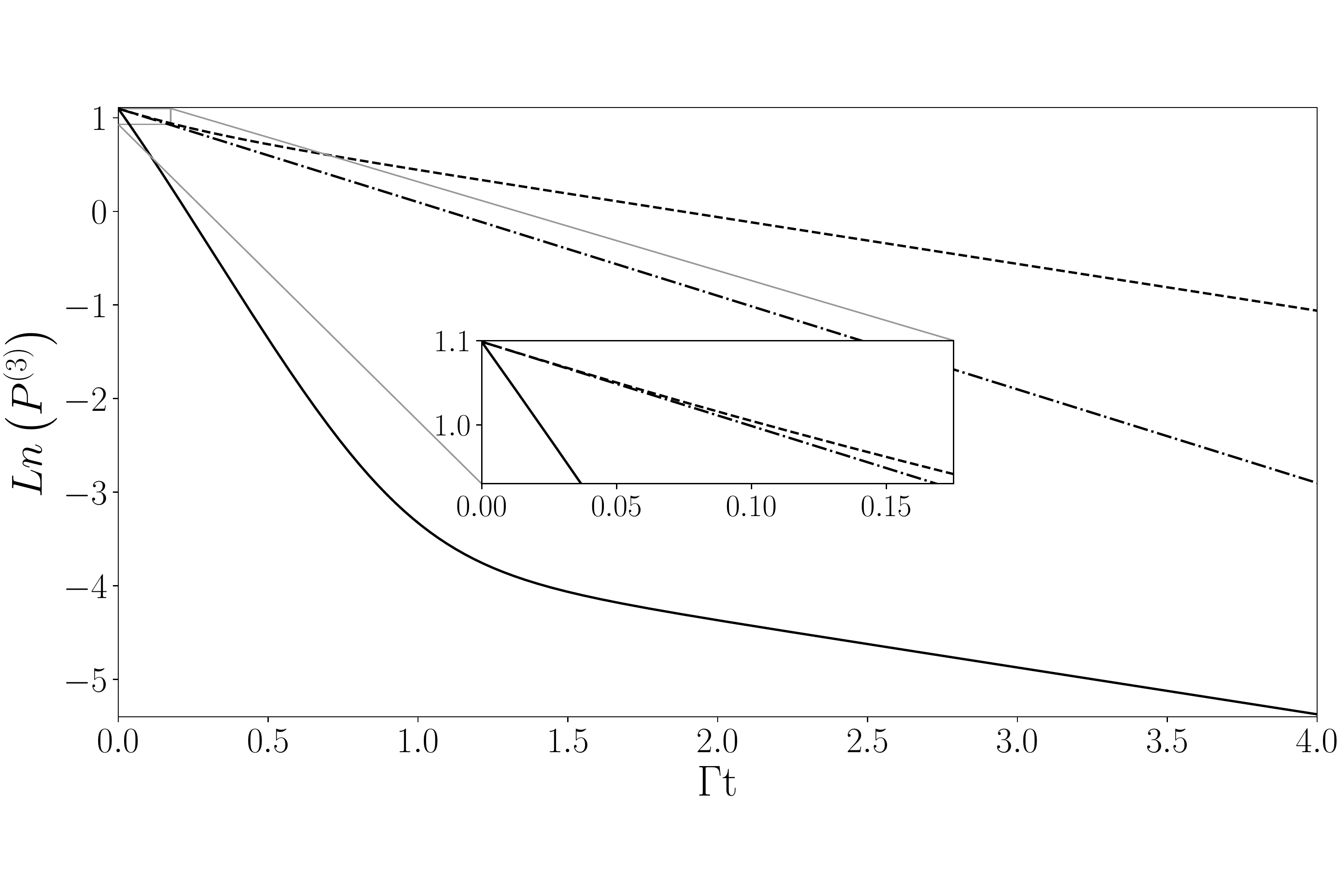}
\caption{Mean number of atomic excitations, in logarithmic scale, as a
  function of time for the symmetric Dicke (solid line) and symmetric
  mixed (dashed line) initial state when $M=3$, $N=10$ and
  $\gamma_c = 0.5\Gamma$. The case of independent emission of atoms,
  $\gamma_c=0$, (dashed-dotted) is plotted as a reference. For long
  times the sub-radiant decay dominates for both cases (symmetric
  Dicke and mixed states). For short times, the initially symmetric
  Dicke state is super-radiant, whereas for the initially symmetric
  mixed state the sub-radiant decay dominates (see inset). }
\label{fig:ir}
\end{figure}

\begin{figure}
\subfloat[Symmetric mixed state\label{sfig:mixed}]{%
  \includegraphics[width=0.485\textwidth]{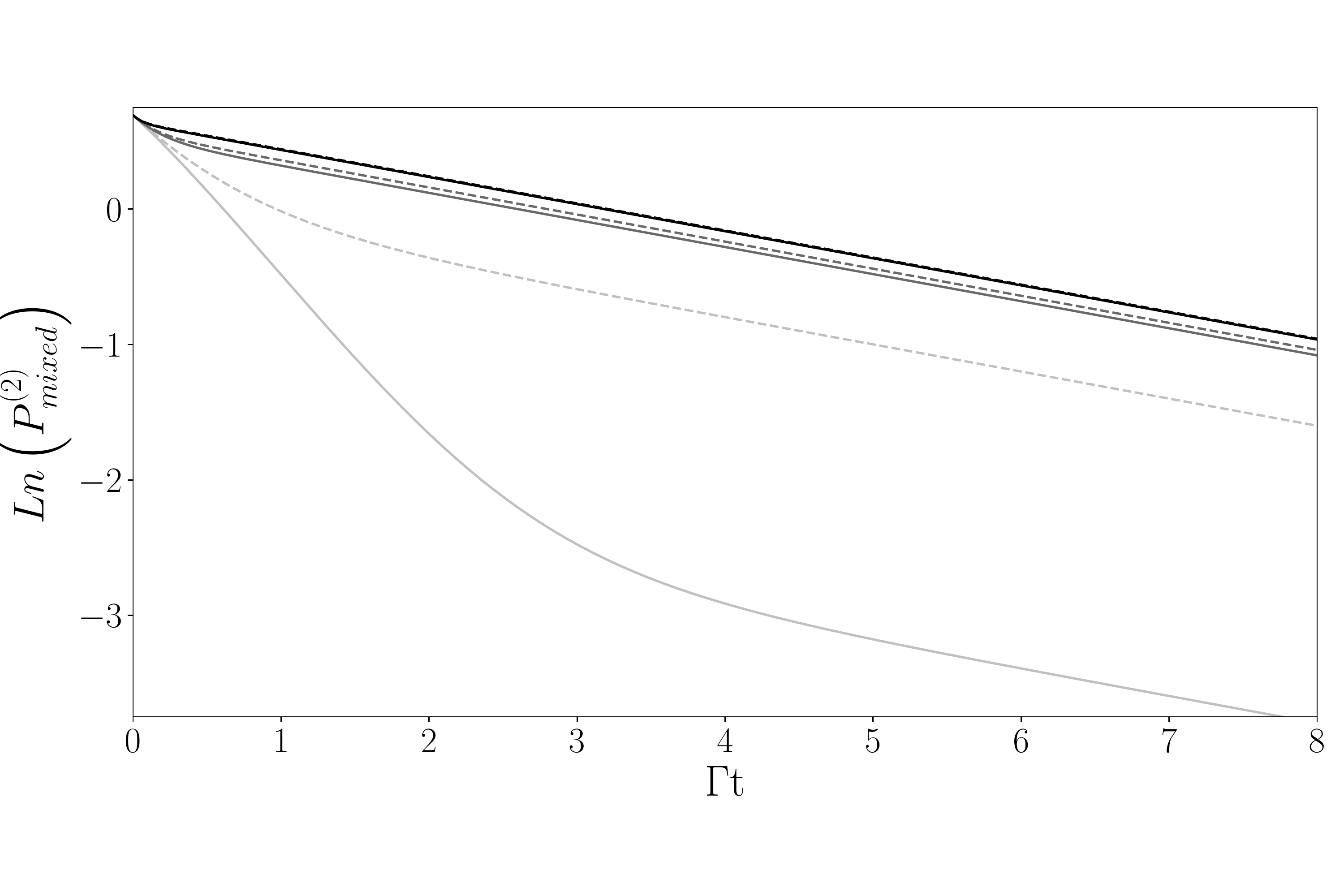}
}\hfill
\subfloat[symmetric Dicke state\label{sfig:dicke}]{%
  \includegraphics[width=0.485\textwidth]{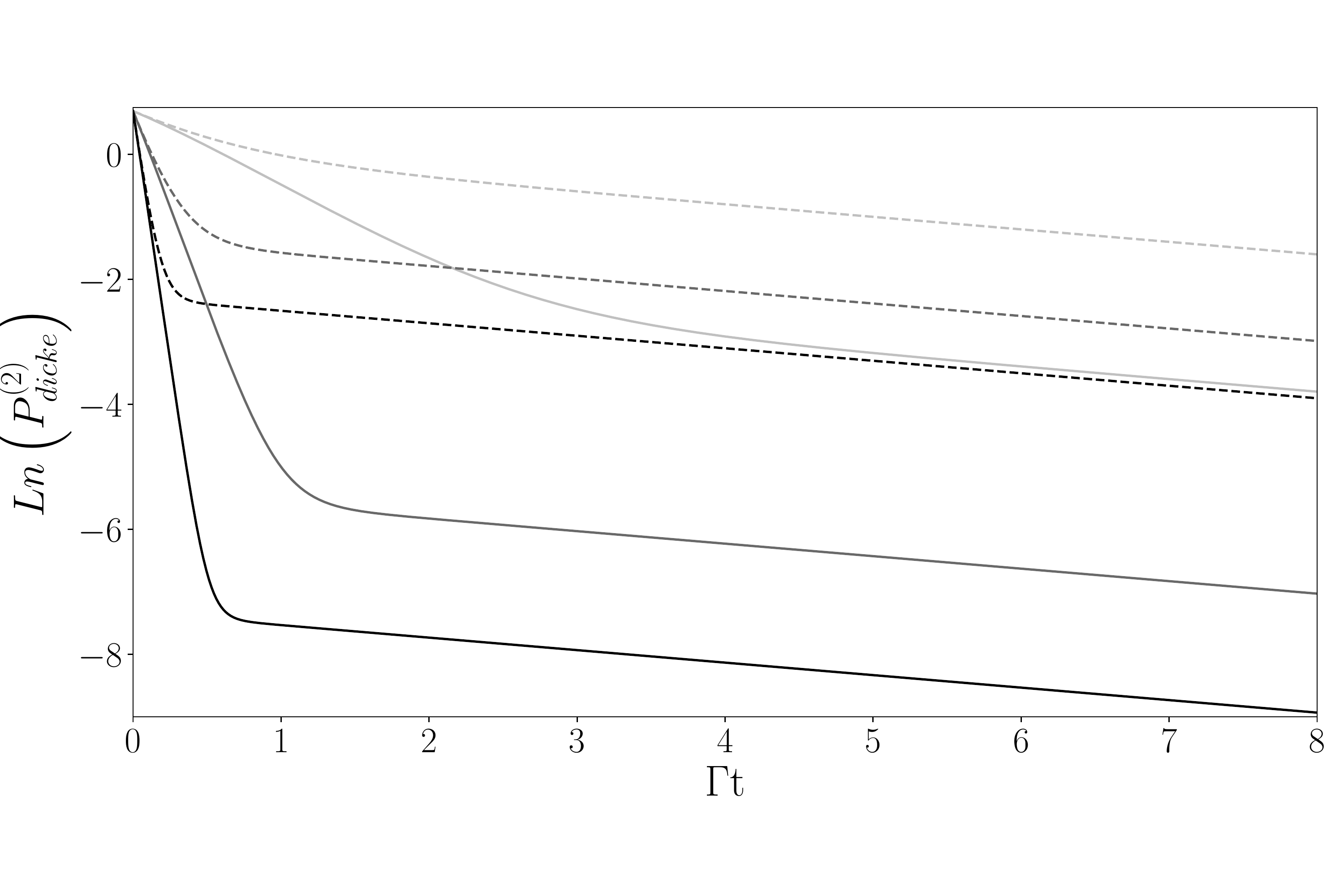}
}
\caption{Mean number of excited atoms, in logarithmic scale, as a
  function of time, calculated using perturbation theory (dashed line)
  and calculated without approximations (solid line). We use
  $\gamma_{c}=0.8\Gamma$ and $M=2$. We plot solutions for $N=2$ (light
  gray line), $N=8$ (dark gray line) and $N=20$ (black line). In (a) we
  compare solutions \eref{ec:irm} and \eref{ir:2mixed}. In (b) we plot
  the expressions \eref{ec:ird} and \eref{ir:2dicke}. Although the
  perturbed solutions were obtained for $\gamma_c \ll \Gamma$, they
  seem to be valid for $N \gg 1$. }
\label{fig:comparison}
\end{figure}

Using \Eref{estado3} we obtain $P^{(M)} (t)=N \alpha_2(t)$. In
\ref{ap5} we show the analytical solution of $P^{(M)}(t)$ for
symmetric Dicke and symmetric mixed states and $M=1,2,3$. For $M = 1$,
\Eref{ir:1dicke} and \Eref{ir:1mixed} coincides with \Eref{ec:irm} and
\Eref{ec:ird} obtained for the perturbative case. When $M = 2, 3$ the
evolution is a sum of decaying exponentials. When $\gamma_c\ll\Gamma$
only two exponential decays are relevant:
$e^{-\left(\Gamma - \gamma_c \right)t}$ and
$e^{-\left[\Gamma + (N-1)\gamma_c \right]t}$. When perturbation theory
is no longer valid, for an initial symmetric Dicke state with $M=2$,
the decay of the number of excitations is a sum of three exponential
terms: the two that appear in perturbation theory, plus
$e^{-2\left[\Gamma + (N-2)\gamma_c \right]t}$. In the case of an
initial symmetric state with $M=2$, the decay of the number of
excitations is the sum of four exponential terms, the three that
appear in the case of an initial symmetric Dicke state plus
$e^{-\left[2\Gamma + (N-4)\gamma_c \right]t}$. In figure~\ref{fig:ir}
we show the mean number of atomic excitations for an initially mixed
symmetric state and for a symmetric Dicke state, when $M=3$, $N=10$
and $\gamma_c=0.5 \, \Gamma$. The evolution shows two clear slopes,
one for short times and one for large times. For both initial
conditions, the sub-radiant behavior dominates for large times. For
short times the evolution of the symmetric Dicke state is
super-radiant. In figure~\ref{sfig:mixed} we compare, for an initial
symmetric mixed state of $M=2$ , the mean number of excited atoms with
(\Eref{ec:irm}) and without (\Eref{ir:2mixed}) perturbation theory,
and for $\gamma_c=0.8\,\Gamma$. In this regime perturbation theory is
no longer justified; nevertheless, when $N\gg 2$ there is no
discernible difference in the figure between the two methods
(perturbation theory and exact results). When $M=N=2$ and $M=2$, $N=8$
the two methods show the same decay at the beginning and sub-radiant
decay at the end the difference between the predictions is the time
where the most sub-radiant decay starts to dominate. In
figure~\ref{sfig:dicke} we repeat the comparison but for an initial
symmetric Dicke state of $M=2$; the difference between the two methods
is the time where the most sub-radiant decay starts to dominate.
Similar results (not shown) are obtained when $M=3,4$.

For all the cases ($M=1,2,3$, initial symmetric Dicke state or symmetric mixed
state) the exponential $e^{-\left(\Gamma - \gamma_c \right)t}$ is the only sub-radiant term in the sum. The sub-radiant decay
always dominates the evolution when $t\gg 1/(N \gamma_c)$. Note that
the sub-radiant decay appears for all the initial conditions if
$\Gamma\neq\gamma_c$ and $M>1$.

We will now study some particular cases.
First we focus on the case where the interaction between the atoms is
maximum, $\Gamma = \gamma_c$. When the initial state is $M=N=2$, we
obtain that the mean number of excited atoms is
\begin{equation}\label{eq:max_sim2}
P^{(2)}(t) = 2 e^{-2 \Gamma  t} (\Gamma  t+1)\, ,
\end{equation}
which coincides with the results in~\cite{dosatomos}. When $M=N=3$ we
obtain
\begin{equation}\label{eq:max_sim3}
P^{(3)}(t) =e^{-3 \Gamma t} (12 \Gamma t-3)+6 e^{-4 \Gamma t}\, .
\end{equation}

When the initial state is the symmetric Dicke state
$\vecop{\rho_{\textrm{dicke}}^{(2)}}$ with $N>2$ (\Eref{eq:initial_dicke}) the
mean number of excited atoms is
\begin{equation}\label{eq:ggcdicke}
P^{(2)}_{\textrm{dicke}}(t) = \frac{2 (N-1) e^{-\Gamma  N t}}{N-2}-\frac{2 e^{-2 \Gamma  (N-1)
    t}}{N-2}\, .
\end{equation}
The time evolution in \Eref{eq:max_sim2}, \Eref{eq:max_sim3} and
\Eref{eq:ggcdicke} have in common that the
number of excited atoms goes to zero when $t\rightarrow\infty$. This observation makes it clear
that in order to have a sub-radiant contribution to the mean number of
atoms excited, it is necessary to take into account atomic independent
decay. When $ \Gamma=\gamma_c$ and the initial condition is
a symmetric Dicke state, the atomic damping basis is not necessary to
obtain analytical results, because the system evolution is closed under
the subspace spanned by the symmetric Dicke states.

The damping basis allow us to obtain results when the initial
state is a symmetric mixed state. For the initial state
$\vecop{\rho_{\textrm{mixed}}^{(2)}}$ with $N>2$ (see
\Eref{eq:initial_mixed}) we obtain
\begin{eqnarray}\label{eq:ggcmixed}
\fl
  P^{(2)}_{\textrm {mixed}}(t) = 2+\frac{2}{N}-\frac{4}{N-1} -\frac{4 e^{-2
                                (N-1)\Gamma   t}}{N (N-1)(N-2)}+
 \frac{2 e^{-(N-2)\Gamma t}}{N}+\frac{4 e^{-N\Gamma   t}}{(N-2)
       N}\, .
\end{eqnarray}
When $t\rightarrow\infty$, the mean number of excited atoms
goes to $2+2/N-4/(N-1)$. The state has a sub-radiant component that
does not exist for the initial pure state $M=N=2,3$ and for an
initial symmetric Dicke state. For $M=3$ we obtain that
\begin{equation}\label{eq:ggcmixedlimit}
\lim_{t\rightarrow\infty}P^{(3)}_{\textrm
  {mixed}}(t) =3+\frac{12}{N-1}-\frac{3}{N}-\frac{12}{N-2}\, .
\end{equation}
When $N\gg 1$ the sub-radiant part is dominant, and in this limit
the system does not decay.

When $\Gamma\neq \gamma_c$ and the initial state is
$\vecop{\rho_{\textrm{mixed}}^{(2)}}$, the weight of the super-radiant
contribution with respect to the sub-radiant contribution in
$P_{\textrm{mixed}}^{(2)} (t)$ increases (compare
\Eref{ir:2mixed} with \Eref{ec:irm}). The reason is that by increasing $\gamma_c$ the coupling between the atoms rises, which
implies that the probability for the system to decay in a super-radiant
state increases.

When $\gamma_c=\Gamma/3$  there are two terms in \Eref{ir:2mixed}
where the denominator is zero. We take the limit
$\gamma_c\rightarrow\Gamma/3$ and obtain
\begin{eqnarray}
\fl
 P_{\textrm{mixed}}^{(2)} (t)= \frac{2 e^{-\frac{1}{3} \Gamma  (N+2) t} [\Gamma t (2N-4) +3 N+6]}{3 N^2}+ 
 \frac{2 \left(N^3-N^2-2 N+2\right)
   e^{-\frac{2 \Gamma t}{3}}}{N^3}-
   \nonumber \\
 \frac{4 e^{-\frac{2}{3} \Gamma  (N+1)
     t}}{N^3}\, .
\end{eqnarray}
The result is a sum of super- and sub- radiant decaying exponentials,
plus a term that consists of an exponential multiplied by time. When
$t\gg 1/(N \Gamma )$ the sub-radiant term, with a rate of
$(2/3)\Gamma$, dominates the evolution.

The operator subspace considered in this section is useful to obtain
the evolution of Dicke and mixed symmetric states defined by Eqs.
\eref{eq:initial_mixed} and \eref{eq:initial_dicke}, but it does not
allow to find the evolution of any symmetric state with at most $M=3$
excitations, for example the superposition of Dicke states with
different excitations. But the method shown in this section can be
applied to any initial state. In order to do so, the subspace that
is closed under the action of operator $\breve{L}$ on the initial
state has to be found. The advantage of the perturbative method is
that, given the maximum number of excitations in the system, the
subspace where the method is going to be used is easily found, as
shown in section~\ref{sc:perturbation}.

\section{Conclusions}\label{sc:conclusions}

The atomic damping basis is a powerful method to study the evolution
of interacting atoms, when the system is symmetric under the
interchange of atomic labels. Using this basis we obtained analytical
expressions for the mean number of atomic excitations for the case of
$M=1,2,3,4$ initially excited atoms, out of a total of $N$ atoms. Our
results, that include the case where the initial state is not pure,
show that the mean number of excited atoms decays as a sum of super-
and sub- radiant exponentials. When there is atomic independent decay
($\Gamma\neq\gamma_c$) and at least two atoms are initially excited,
the sub-radiant component of the evolution always appears in the
solutions that we studied, and dominates the system evolution for large
times.

\ack
We thank David P. Sanders for proofreading the manuscript. This work
was supported by DGAPA-UNAM under grant PAPIIT-IG120518.

\appendix
\section{Some linear maps in terms of collective superoperators}{\label{ap1}}
The following are some useful identities between superoperators:

\[
\fl
\begin{array}{ccc}
\sum_\mu \hat{\sigma}_{-}^{(\mu)} \vecop{\rho} \hat{\sigma}_{+}^{(\mu)} = \supop{A}{00}{11}_+ \vecop{\rho} 
&\qquad & \sum_\mu \hat{\sigma}_{+}^{(\mu)} \vecop{\rho} \hat{\sigma}_{-}^{(\mu)} = \supop{A}{11}{00}_+ \vecop{\hat{\rho}}
\\
\sum_\mu \hat{\sigma}_{+}^{(\mu)} \hat{\sigma}_{-}^{(\mu)}\vecop{\rho}  = 
\left( \supop{A}{10}{10}_+ + \supop{A}{11}{11}_+ \right) \vecop{\rho}
& \qquad &
\sum_\mu  \vecop{\rho} \hat{\sigma}_{+}^{(\mu)} \hat{\sigma}_{-}^{(\mu)} = 
\left( \supop{A}{01}{01}_+ + \supop{A}{11}{11}_+ \right) \vecop{\rho}
\\
\sum_\mu \hat{\sigma}_{-}^{(\mu)} \hat{\sigma}_{+}^{(\mu)}\vecop{\rho}  =  
\left( \supop{A}{01}{01}_+ + \supop{A}{00}{00}_+ \right) \vecop{\rho}
& \qquad &
\sum_\mu \vecop{\rho} \hat{\sigma}_{-}^{(\mu)} \hat{\sigma}_{+}^{(\mu)} = 
\left( \supop{A}{10}{10}_+ + \supop{A}{00}{00}_+ \right) \vecop{\rho} 
\\
\hat{J}_+ \vecop{\rho} =
\left( \supop{A}{10}{00}_+ + \supop{A}{11}{01}_+ \right) \vecop{\rho}
&\qquad &
\vecop{\rho}\hat{J}_+ = 
\left( \supop{A}{10}{11}_+ + \supop{A}{00}{01}_+ \right) \vecop{\rho}
\\
\hat{J}_- \vecop{\rho} = 
\left( \supop{A}{00}{10}_+ + \supop{A}{01}{11}_+ \right) \vecop{\rho}
& \qquad &
\vecop{\rho}\hat{J}_- = 
\left( \supop{A}{11}{10}_+ + \supop{A}{01}{00}_+ \right) \vecop{\rho}
\end{array}\]
\[
\fl
\begin{array}{c}
\hat{J}_- \vecop{\rho} \hat{J}_+ = \left( \supop{A}{00}{10}_+ + \supop{A}{01}{11}_+ \right)
\left( \supop{A}{10}{11}_+ + \supop{A}{00}{01}_+ \right) \vecop{\rho}
\\
\hat{J}_+ \vecop{\rho} \hat{J}_- = 
\left( \supop{A}{10}{00}_+ + \supop{A}{11}{01}_+ \right)
\left( \supop{A}{11}{10}_+ + \supop{A}{01}{00}_+ \right) \vecop{\rho}
\end{array}\]

\section{Eigenvectors of $\breve{L}_{i}$ for 3 excitations.}
\label{ap2}
We show the basis for a subspace of $N$ atoms with at most
three excitations. In this approximation we consider zero all those
eigenvectors with less than $N-3$ atoms in the state ground. The right
eigenvectors of $\breve{L}_{i}$ ($\gamma_{10}=\Gamma-\gamma_c$) are
{\scriptsize
\begin{eqnarray}
\fl
\lambda_{N-3,0} = -\frac{3\gamma_{10}}{2}: \vecop{N-3,0}_0 = \frac{N(N-1)(N-2)}{6}\lind{Q}{N&0}{0&0}, \quad \vecop{N-3,0}_1 = \frac{N(N-1)(N-2)}{2}\lind{Q}{N-3&2}{1&0}, \nonumber\\
\vecop{N-3,0}_2 = \frac{N(N-1)(N-2)}{2} \lind{Q}{N-3&1}{2&0},
\quad \vecop{N-3,0}_3= \frac{N(N-1)(N-2)}{6}\lind{Q}{N-3&0}{3&0}
\nonumber\\
\fl
\lambda_{N-2,0} = -\gamma_{10}: \vecop{N-2,0}_0 = \frac{N(N-1)}{3}\lind{Q}{N-2&2}{0&0}, \quad \vecop{N-2,0}_1 = N(N-1)\lind{Q}{N-2&1}{1&0}, 
\nonumber \\ \vecop{N-2,0}_2 =
\frac{N(N-1)}{2}\lind{Q}{N-2&0}{2&0}
\nonumber\\
\fl
\lambda_{N-2,1} = -2\gamma_{10}: \vecop{N-2,1}_0 = -\frac{N(N-1)(N-2)}{2} \left(\lind{Q}{N-2&2}{0&0} - \lind{Q}{N-3&2}{0&1}\right), 
\nonumber\\
\vecop{N-2,1}_1 = -N(N-1)(N-2) \left( \lind{Q}{N-2&1}{1&0} - \lind{Q}{N-3&1}{1&1} \right), 
\nonumber\\
\vecop{N-2,1}_2 = - \frac{N(N-1)(N-2)}{2} \left( \lind{Q}{N-2&0}{2&0} - \lind{Q}{N-3&0}{2&1} \right)
\nonumber\\
\fl
\lambda_{N-1,0} = -\frac{\gamma_{10}}{2}: \vecop{N-1,0}_0 = N \lind{Q}{N-1&1}{0&0}, \qquad \vecop{N-1,0}_1 = N \lind{Q}{N-1&0}{1&0}
\nonumber\\
\fl
\lambda_{N-1,1} = -\frac{3\gamma_{10}}{2}: \vecop{N-1,1}_0 = -N(N-1) \left( \lind{Q}{N-1&1}{0&0} - \lind{Q}{N-2&1}{0&1} \right),
\nonumber \\ 
\vecop{N-1,1}_1 = -N(N-1) \left( \lind{Q}{N-1&0}{1&0} - \lind{Q}{N-2&0}{1&1} \right)
\nonumber\\
\fl
\lambda_{N-1,2} = -\frac{5\gamma_{10}}{2}: \vecop{N-1,2}_0 = \frac{N(N-1)(N-2)}{2} \left( \lind{Q}{N-1&1}{0&0} - 2\lind{Q}{N-2&1}{0&1} + \lind{Q}{N-3&1}{0&2} \right), 
\nonumber\\
\vecop{N-1,2}_1 = \frac{N(N-1)(N-2)}{2} \left( \lind{Q}{N-1&0}{1&0} -2\lind{Q}{N-2&0}{1&1} + \lind{Q}{N-3&0}{1&2}
\right) 
\nonumber\\
\fl
\lambda_{N,0} = 0: \vecop{N,0}_0 = \lind{Q}{N&0}{0&0}
\nonumber\\
\fl
\lambda_{N,1} = -\gamma_{10}: \vecop{N,1}_0 = -N \left( \lind{Q}{N&0}{0&0} - \lind{Q}{N-1&0}{0&1} \right)
\nonumber\\
\fl
\lambda_{N,2} = -2\gamma_{10}: \vecop{N,2}_0 =  \frac{N(N-1)}{2} \left( \lind{Q}{N&0}{0&0} -2\lind{Q}{N-1&0}{0&1} + \lind{Q}{N-2&0}{0&2}
\right)
\nonumber\\
\fl
\lambda_{N,3} = -3\gamma_{10}: \vecop{N,3}_0 = - \frac{N(N-1)(N-2)}{6} \left(\lind{Q}{N&0}{0&0} - 3\lind{Q}{N-1&0}{0&1} + 3\lind{Q}{N-2&0}{0&2} - \lind{Q}{N-3&0}{0&3} \right)
\nonumber
\end{eqnarray}}

If we perturb the operator $\breve{L}_i$ with $\breve{L}_{c}$ we
obtain, to first order in the eigenvalues $\Lambda_{n,m}$, and
zero order in the right eigenvectors $\vecop{\phi_{n,m}}$, the following
results: {\tiny
\[
\fl
\begin{array}{cccc}
\Lambda_{0,1} = 0: &
\vecop{\phi_{0,1}} = \vecop{N,0}_0 
 & 
\Lambda_{3,4} = -\frac{3}{2}\gamma_{10} \left( 1 + \frac{N-2}{3} \gamma_c \right): & \vecop{\phi_{3, 4}} = -\frac{2}{N-2}\vecop{N-3,0}_1 + \vecop{N-1,1}_0 
\\
\Lambda_{1,1} = -\frac{1}{2}\gamma_{10}\big( 1 + N\gamma_c\big):&
\vecop{\phi_{1,1}} = \vecop{N-1,0}_0 
 & 
\Lambda_{3,5} = -\frac{3}{2}\gamma_{10} \left( 1 + \frac{3N-2}{3} \gamma_c \right): & \vecop{\phi_{3, 5}} = \vecop{N-3,0}_2 + \vecop{N-1,1}_1 
\\
\Lambda_{1,2} = -\frac{1}{2}\gamma_{10}\big( 1 + N\gamma_c\big):& \vecop{\phi_{1,2}} = \vecop{N-1,0}_1
 & 
\Lambda_{3,6} = -\frac{3}{2}\gamma_{10} \left( 1 + \frac{N-2}{3} \gamma_c \right): & \vecop{\phi_{3, 6}} = -\frac{2}{N-2}\vecop{N-3,0}_2 + \vecop{N-1,1}_1 
 \\
\Lambda_{2,1} = -\gamma_{10}\big[ 1 + (N-1)\gamma_c\big]: &
\vecop{\phi_{2,1}} = \vecop{N-2,0}_0 
 &
 \Lambda_{4,1} = -2\gamma_{10} \left( 1 + \frac{N}{2} \gamma_c \right): & \vecop{\phi_{4, 1}} = \vecop{N-2,1}_0
\\
\Lambda_{2,2} = -\gamma_{10}\big[ 1 + (N-1)\gamma_c\big]: &
\vecop{\phi_{2,2}} = \vecop{N-2,0}_2 
 & 
\Lambda_{4,2} = -2\gamma_{10} \left( 1 + \frac{N}{2} \gamma_c \right): &
\vecop{\phi_{4, 2}} =\vecop{N-2,1}_2 
\\
\Lambda_{2,3} = -\gamma_{10} \left( 1 + N \gamma_c \right): &
\vecop{\phi_{2, 3}} = \vecop{N-2,0}_1 + \vecop{N,1}_0 
 & 
\Lambda_{4,3} = -2\gamma_{10} \left( 1 + \frac{N+2+\nu}{4} \gamma_c \right): &
\vecop{\phi_{4, 3} } = \frac{N-2+\nu}{4N-8}\vecop{N-2,1}_1 + \vecop{N,2}_0 
\\
\Lambda_{2,4} = -\gamma_{10}:&
\vecop{\phi_{2, 4}} =-\frac{1}{N-1}\vecop{N-2,0}_1 + \vecop{N,1}_0
 & 
\Lambda_{4,4} = -2\gamma_{10} \left( 1 + \frac{N+2-\nu}{2} \gamma_c \right): &
\vecop{\phi_{4,4}} = \frac{N-2-\nu}{4N-8}\vecop{N-2,1}_1 + \vecop{N,2}_0
\\
\Lambda_{3,1} = -\frac{3}{2} \gamma_{10}\big[ 1 + (N-2)\gamma_c\big]: & \vecop{\phi_{3,1}} = \vecop{N-3,0}_0 
 & 
\Lambda_{5,1} = -\frac{5}{2}\gamma_{10}\big( 1 + \frac{N+4}{5}\gamma_c\big):&
\vecop{\phi_{5,1}} = \vecop{N-1,2}_0
\\ 
	\Lambda_{3,2} = -\frac{3}{2} \gamma_{10}\big[ 1 + (N-2)\gamma_c\big]: & \vecop{\phi_{3,2}} = \vecop{N-3,0}_3 
 &
 \Lambda_{5,2} = -\frac{5}{2}\gamma_{10}\big( 1 + \frac{N+4}{5}\gamma_c \big): &
\vecop{\phi_{5,2}} = \vecop{N-1,2}_1
	\\ 
\Lambda_{3,3} = -\frac{3}{2}\gamma_{10} \left( 1 + \frac{3N-2}{3} \gamma_c \right): & \vecop{\phi_{3, 3}} = \vecop{N-3,0}_1 + \vecop{N-1,1}_0 
 & 
\Lambda_{6,1} = -3\gamma_{10}\big( 1 + \gamma_c\big): &
\vecop{\phi_{6,1}} = \vecop{N,3}_0 
\end{array}\]} 
\noindent with $\nu = \sqrt{(N+6)(N-2)}$. Note that the eigenvalues remain
degenerate at first order.

With this basis we can solve \Eref{eq:master_equation} for
states \eref{eq:initial_mixed} and \eref{eq:initial_dicke}. As an
example, for $M=3$ we get
{\small
\begin{eqnarray}
\fl
\vecop{\rho^{(3)}(t)} = \vecop{\phi_{0,1}}
+ \frac{3 k_1\, e^{-\left( \gamma_{10} + N \gamma_c \right) t}}{N^2} \vecop{\phi_{2,3}}
+ \frac{3 k_2 \, e^{-\gamma_{10}t}}{N^2} \vecop{\phi_{2,4}}
 +
\frac{ 3k_3 \, e^{-2 \left(\gamma_{10} + \frac{N+2+\sqrt{(N+6)(N-2)}}{4}\gamma_c \right) t}}{N(N-1)}
\vecop{\phi_{4,3}}
\nonumber \\ 
 +
\frac{ 3k_4 \, e^{-2 \left(\gamma_{10} + \frac{N+2-\sqrt{(N+6)(N-2)}}{4}\gamma_c \right) t}}{N(N-1)}
\vecop{\phi_{4,4}}
 + 
\frac{6 e^{-3 \left(\gamma_{10} + \gamma_c \right)t}}{N(N-1)(N-2)}
               \vecop{\phi_{6,1}}\, ,
\label{soluciontresexc}
\end{eqnarray}}
where for mixed states we have
\begin{eqnarray}
k_1 &=& 1,\qquad k_2=N-1, \qquad  
k_3 = \frac{N+6-\sqrt{(N+6)(N-2)}}{N+6}, 
\nonumber \\
k_4 &=& \frac{N+6+\sqrt{(N+6)(N-2)}}{N+6}\, ,
\label{soluciontresexc-mixed}
\end{eqnarray}
and for symmetric Dicke states the coefficients are
\begin{eqnarray}
k_1 &=& N-2, \qquad k_2=2, \qquad
k_3 = \frac{3N-10+\sqrt{(N+6)(N-2)}}{\sqrt{(N+6)(N-2)}},
\nonumber \\
k_4 &=& - \frac{3N-10-\sqrt{(N+6)(N-2)}}{\sqrt{(N+6)(N-2)}}\, .
\label{soluciontresexc-dicke}
\end{eqnarray}
We observe sub- and super-radiant terms due to interaction between the
atoms. 

\section{Damping basis without perturbation for a system with three excitations.}\label{ap3}
If we consider three excited atoms among a total of $N$, the linear
map $\breve{L}_{i}+\breve{L}_c$ is closed for the ordered set
$\mathcal{A}_3 = \{\vecop{N,0}_0, \vecop{N,1}_0, \vecop{N-2,0}_1,
\vecop{N,2}_0, \vecop{N-2,1}_1,$
$\vecop{N-4,0}_2, \vecop{N,3}_0, \vecop{N-2,2}_1,\vecop{N-4,1}_2,
\vecop{N-6,0}_3 \}$. The matrix form of the superoperator in this
subspace is

{\tiny
\[
\setlength{\arraycolsep}{1pt}
\renewcommand{\arraystretch}{0.8}
\fl
-
\left(
\begin{array}{cccccccccc}
0& 0 & 0 & 0 & 0 & 0 & 0 & 0 & 0 & 0\\
0& \gamma_{10}+\gamma_c & (N-1)\gamma_c & 0 & 0 & 0 & 0 &0 &0 &0 \\
0& \gamma_c & \gamma_{10}+(N-1)\gamma_c & - 2\gamma_c & -2(N-2)\gamma_c & 0 &0 &0 &0 &0\\
0& 0 & 0 & 2\gamma_{10}+2\gamma_c & 2(N-2)\gamma_c & 0 &0 &0 &0 &0 \\ 
0& 0 & 0 & \gamma_c & 2\gamma_{10}+N\gamma_c & (N-3)\gamma_c & -2\gamma_c & -2(N-3)\gamma_c &0 &0 \\
0& 0 & 0 & 0 & 4\gamma_c & 2\gamma_{10}+2(N-3)\gamma_c &0 & - 8\gamma_c &- 4(N-4)\gamma_c &0 \\
0& 0 &0 &0 &0 &0 &3\gamma_{10} + 3\gamma_c &3(N-3)\gamma_c &0 &0 \\ 
0& 0 &0 &0 &0 &0 &\gamma_c &3\gamma_{10}+(N+1)\gamma_c &2(N-4)\gamma_c &0\\ 
0& 0 &0 &0 &0 &0 &0 &4\gamma_c &3\gamma_{10}+(2N-5)\gamma_c &(N-5)\gamma_c\\
0& 0 &0 &0 &0 &0 &0 &0 &9\gamma_c &3\gamma_{10}+3(N-5)\gamma_c   
\end{array} \right)
\]}

By diagonalizing this matrix we get analytic expressions for the
eigenvectors and the eigenvalues that allow us to obtain the evolution
of the states \eref{eq:initial_mixed} and \eref{eq:initial_dicke}
without perturbation and for $M=1,\, 2, \, 3$.

\section{Population of excited atoms}\label{ap5}
The mean number of excited atoms $P^{(M)}$, as a function of time, is
shown below for $M = 1, 2, 3$ excitations, and for initially symmetric
mixed states \eref{eq:initial_mixed} and symmetric Dicke states
\eref{eq:initial_dicke}. Using $\gamma_{10}=\Gamma-\gamma_c$ we obtain:
\begin{eqnarray}
\fl
P_{\textrm{dicke}}^{(1)}(t) = e^{-\left[\Gamma + (N-1)\gamma_c \right]t}\, ,
 \label{ir:1dicke}
\end{eqnarray}

\begin{eqnarray}
\fl
P_{\textrm{mixed}}^{(1)}(t)=\frac{N-1}{N}e^{-\left( \Gamma - \gamma_c \right) t} + \frac{1}{N} e^{-\left[ \Gamma + (N-1)\gamma_c \right] t}\, ,
 \label{ir:1mixed}
\end{eqnarray}

{\footnotesize
\begin{eqnarray}
\fl
P_{\textrm{dicke}}^{(2)}(t) = \frac{2}{N \left[ \gamma _{10} + (N-2) \gamma _c\right] \left[\gamma _{10} + 2 (N-1) \gamma _c\right]} \bigg[
\left \{ (N-2)\gamma_{10} \gamma_c + \gamma_{10}^2 \right \} e^{- \left(\Gamma - \gamma_c \right) t} + \bigg\{\gamma _{10} \gamma _c  \left(3 N^2-5 N+2 \right)
\nonumber 
\\ 
+ (N-1)\gamma_{10}^2 + 2N(N-1)^2 \gamma_c ^2 \bigg \} e^{ -\left[\Gamma + (N-1) \gamma_c \right]t}
- 2N (N-1)  \gamma _c^2 e^{-2\left[\Gamma + (N-2) \gamma _c\right] t} \bigg] \, ,
 \label{ir:2dicke}
\end{eqnarray}}

{\scriptsize
\begin{eqnarray}
\fl
P_{\textrm{mixed}}^{(2)} (t) = 2\Bigg[
\left(
\frac{N-1}{N} - 
\frac{ 2 \gamma_c \left[\gamma_{10} + 2 (N-2) \gamma _c\right]}{N\left[\gamma_{10} + (N-2)\gamma_c \right]\left[\gamma _{10} + 2 (N-1) \gamma _c \right]}
 \right)
 e^{-\left(\Gamma - \gamma_c \right)t} + 
\left(\frac{1}{N}
+ \frac{2\gamma_c \left[\gamma _{10} + (N-4) \gamma _c \right]}{N\left[\gamma _{10}-2 \gamma _c\right]\left[\gamma_{10} + (N-2)\gamma_c \right] } \right) e^{- \left[\Gamma + (N-1)  \gamma _c \right] t} 
\nonumber\\ 
- \frac{4 \gamma _c^2 e^{-2\left[\Gamma + (N-2) \gamma _c \right]t}}{N\left[\gamma_{10} + (N-2)\gamma_c \right] \left[\gamma _{10} + 2 (N-1) \gamma _c\right]} - \frac{2(N-2) \gamma _c^2 e^{- \left[2 \Gamma + (N-4) \gamma _c\right]t}}{N\left[\gamma _{10}-2 \gamma _c\right] \left[\gamma_{10} + (N-2)\gamma_c \right] } \Bigg]\, ,
 \label{ir:2mixed}
\end{eqnarray}
}

{\scriptsize
\begin{eqnarray}
\fl
P_{\textrm{dicke}}^{(3)} (t)= 3  \Bigg[\frac{4 (N-2) \gamma _c^3 \left[\gamma _{10} + 3 (N-2) \gamma _c\right] e^{ -\left[ 3 \Gamma +  (3N-7) \gamma _c \right]}}{\left[\gamma _{10}+(N-4) \gamma _c\right] \left[\gamma _{10}+(N-3) \gamma _c\right] \left[\gamma _{10}+2 (N-2) \gamma _c\right] \left[2 \gamma _{10} + 3 (N-2) \gamma _c\right]}-
\frac{4 \gamma _{10} (N-2) \gamma _c^2 e^{- \left[ 2\Gamma + (N-4) \gamma _c \right]t}}{N \left[\gamma _{10}-2 \gamma _c\right] \left[\gamma _{10} + (N-2) \gamma _c\right] \left[\gamma _{10} + 2 (N-2) \gamma _c\right]} 
\nonumber \\ \fl
- \frac{4 (N-2) (N-1) \left(\gamma _{10} + N \gamma _c\right)
\gamma _c^2  e^{ -2\left[  \Gamma + (N-2) \gamma _c \right] t}}{N \left[\gamma _{10} + (N-4) \gamma _c\right] \left[\gamma _{10} + (N-2) \gamma _c\right] \left[\gamma _{10} + 2 (N-1) \gamma _c\right]}+
 \frac{2 \gamma _{10} \left( 2 \gamma _{10}^2 + (5 N-6)\gamma _{10} \gamma _c+(N-2) (3N-4) \gamma _c^2\right) e^{-\left(\Gamma - \gamma _c \right)t}}{N \left[\gamma _{10} +
(N-2) \gamma _c \right] \left[2 \gamma _{10} + 3 (N-2) \gamma _c\right] \left[\gamma _{10}
+ 2 (N-1) \gamma _c\right]}+
\nonumber \\ \fl
\frac{(N-2) \left(\gamma _{10}^3 + (2N-3) \gamma _{10}^2  \gamma _c+ (N-4) (N-1) \gamma _{10} \gamma _c^2-2 N(N-1) \gamma _c^3\right) e^{- \left[\Gamma + (N-1) \gamma _c \right]t}}{N \left[\gamma _{10}-2 \gamma _c\right] \left[\gamma _{10} + (N-3) \gamma _c \right] \left[\gamma _{10} + (N-2) \gamma _c \right]}\Bigg]\, ,
 \label{ir:3dicke}
\end{eqnarray}}

{\scriptsize
\begin{eqnarray}
\fl
P_{\textrm{mixed}}^{(3)} = 3 \Bigg[\frac{4(N-4)(N-3) \gamma _c^3 e^{- \left[3 \Gamma +(N-7) \gamma _c\right] t}}{(N-2) (N-1) \left[\gamma _{10}-2 \gamma _c\right]^2 \left[2 \gamma _{10}+(N-4) \gamma _c\right]}+
\frac{16 (N-3) \left[\gamma _{10}+(N-6) \gamma _c\right] \gamma _c^3 e^{ - \left[3 \Gamma  + (2N-9) \gamma _c\right]t}}{N(N-2) \left[\gamma _{10}-4 \gamma _c\right] \left[2 \gamma _{10}+(N-6) \gamma _c\right] \left[\gamma _{10}+(N-4) \gamma _c\right] \left[\gamma _{10}+(N-3) \gamma _c\right]} 
\nonumber \\ \fl
+ \frac{24 \left[\gamma _{10}+3 (N-2) \gamma _c\right] \gamma _c^3 e^{- 3 \left[\Gamma +(N-3) \gamma _c\right] t} }{N(N-1)\left[\gamma _{10}+(N-4) \gamma _c\right] \left[\gamma _{10}+(N-3) \gamma _c\right] \left[\gamma _{10}+2 (N-2) \gamma _c\right] \left[2 \gamma _{10}+3 (N-2) \gamma _c\right]} -
\nonumber \\ \fl
\frac{8  \left[\gamma _{10}^2+(N-4) \gamma _c \gamma _{10}-12 \gamma _c^2\right] \gamma _c^2 e^{-2 \left[\Gamma +(N-2) \gamma _c\right] t}}{N \left[\gamma _{10}-4 \gamma _c\right] \left[\gamma _{10}+(N-4) \gamma _c\right] \left[\gamma _{10}+(N-2) \gamma _c\right] \left[\gamma _{10}+2 (N-1) \gamma _c\right]} -
\nonumber \\  \fl
\frac{4  \left[(N-2) \gamma _{10}^3+[12 + N(3 N-14) ] \gamma _c \gamma _{10}^2 + 2 \{N [14 + N (N-8)]+4\} \gamma _c^2 \gamma _{10} - 8 (N-3) (N-2) \gamma _c^3 \right] \gamma _c^2 e^{- \left[2 \Gamma + (N-4) \gamma _c \right] t}}{N \left[\gamma _{10}-2 \gamma _c\right]^2 \left[\gamma _{10}+(N-4) \gamma _c\right] \left[\gamma _{10}+(N-2) \gamma _c\right] \left[\gamma _{10}+2 (N-2) \gamma _c\right]} + 
\nonumber \\ \fl
\frac{\left[2 \gamma _{10}^5+(5N-16) \gamma _c \gamma _{10}^4+[26 + N(4N-27)] \gamma _c^2 \gamma _{10}^3 + \{N [12 + N(N-11)]+44\} \gamma _c^3 \gamma _{10}^2 - 8 [3 + N(N-7) ] \gamma _c^4 \gamma _{10}+24 (N-6) \gamma _c^5\right] }{N \left[\gamma _{10}-2 \gamma _c\right]^2 \left[2 \gamma _{10}+(N-6) \gamma _c\right] \left[\gamma _{10}+(N-3) \gamma _c\right] \left[\gamma _{10}+(N-2) \gamma _c\right]} 
\nonumber \\ \fl
\times e^{- \left[\Gamma + (N-1) \gamma _c\right] t} + \frac{e^{- \left( \Gamma -\gamma_c \right) t }}{N \left[2 \gamma _{10}+(N-4) \gamma _c\right] \left[\gamma _{10}+(N-3) \gamma _c\right] \left[\gamma _{10}+(N-2) \gamma _c\right] \left[2 \gamma _{10}+3 (N-2) \gamma _c\right] \left[\gamma _{10}+2 (N-1) \gamma _c\right]}
\Bigg( 4 (N-1) \gamma _{10}^5
\nonumber \\ \fl
+8 [3N (N-3)+4] \gamma _c \gamma _{10}^4 + 
\{N [5 N (11N-57)+378]-52\} \gamma _c^2 \gamma _{10}^3 
+  6 (N-4) (N-3) (N-2) \left(N^3 -4 N^2+N-2\right) \gamma _c^5  +
\nonumber \\ \fl
(N \{3 N [N (20 N-151)+357]-730\}-88) \gamma _c^3 \gamma _{10}^2 
+(N-3) (N \{N [N (31 N-221)+446]-192\}
-16) \gamma _c^4 \gamma _{10}  \Bigg)
\Bigg]\, .
 \label{ir:3mixed}
\end{eqnarray}}

\section*{References}
\bibliography{referencias}

\end{document}